\documentclass[showpacs,amssymb,twocolumn,aps]{revtex4}
\usepackage{amssymb}
\usepackage{amsmath}
\usepackage[pdftex]{graphicx}
\usepackage{epstopdf}
\usepackage[utf8]{inputenc}
\usepackage{fancyhdr}
\usepackage{slashed}
\usepackage{turnstile}

\begin{document}

\date{\today}
\title{High temperature limit in static backgrounds}

\author{F. T. Brandt  \footnote{fbrandt@usp.br}
and J. B. Siqueira \footnote{joao@fma.if.usp.br} }
\affiliation{ Instituto de F\'{\i}sica,
Universidade de S\~ao Paulo,
S\~ao Paulo, SP 05315-970, Brazil}

\begin{abstract}
We prove that the hard thermal loop contribution to static thermal amplitudes can be obtained by setting all the external four-momenta to zero
before performing the Matsubara sums and loop integrals. At the one-loop order we do an iterative procedure for all the 1PI (one-particle irreducible) one-loop diagrams and at
the two-loop order we consider the self-energy. Our approach is sufficiently general to the extent that it includes theories with
any kind of interaction vertices, such as gravity in the weak field approximation, for $d$ space-time dimensions. This result is valid whenever the external
fields are all bosonic. 
\end{abstract}


\maketitle

\section{Introduction}
The high-temperature limit of the one-loop thermal Green's functions enjoys some important properties which allow us to  obtain a closed-form expression for the corresponding effective actions 
of Abelian as well as non-Abelian gauge theories. One such property is the existence of simple Ward identities, which relate the $n$-point with the ($n+1$)-point functions, reflecting the underlying
gauge invariance of the effective action. Also, these high-temperature Green's functions have a characteristic analytic dependence
on the external momenta, 
yielding effective actions which are nonlocal functionals of the external
fields \cite{Braaten:1990mz,Frenkel:1990br}. In the high-temperature limit, it is also possible
to sum an infinite set of higher loop contributions to the free energy, the so-called ring diagrams, which are individually 
infrared divergent \cite{kapusta:book89,lebellac:book96,Brandt:2009rq,Brandt:2006bf}. Similar properties are also 
encountered in quantum gravity, where the Ward identities emerge as a consequence of the invariance under
general coordinate transformations \cite{Brandt:1993dk,Brandt:1995mv}. However, in the case of gravity, it has not been possible to find an explicit closed-form 
expression for the underlying, nonlocal, effective action as a functional of the metric field.

For the special case when the metric fields are static, we have shown recently that it is possible to sum all the 1PI one-loop $n$-graviton
functions in terms of a local closed-form expression for the effective action in the high-temperature limit \cite{Brandt:2012ei}. 
A key ingredient in order to obtain this result was the use of the identity between the high-temperature static and 
the zero four-momentum limit of all the one-loop thermal Green's functions. Throughout this work,
the equivalence between the {\it static} hard thermal loop and the {\it zero-momentum} thermal amplitudes will be abbreviated as {\it SZM identity}.
This property has been shown to be true 
by explicitly computing the two- and three-point functions at one-loop order \cite{Frenkel:2009pi}.
On the other hand, in the limit where all the spacial components of the external momenta vanish (long-wavelength limit),
the thermal Green's functions are not the same as in the static limit
for high-temperatures \cite{Brandt:2009ht}.
For example the high-temperature limit of the self-energy has different static- and long wavelength limits \cite{lebellac:book96}.
This occurs because in the long-wavelength limit the analytic continuation from discrete to continuous external energies would modify 
the integrands of the thermal Green's functions in a nontrivial way. (Throughout this work, we will employ the imaginary time formalism \cite{lebellac:book96,kapusta:book89}.) 
Another way to understand why the long-wavelength and the high-temperature limits do not commute is to notice that
the leading high-temperature contribution arises from the region where the loop three-momentum  is much larger than
any external three-momentum \cite{Braaten:1990mz}.

The static limit of the hard thermal loop 
contributions to thermal amplitudes can be quite cumbersome in general, specially when the amplitudes have more than two external legs. This is because one usually would have to keep the external three-momenta 
nonzero before the integration over the loop momenta is performed; only in the end of the calculation would the external three-momenta be reduced to zero. 
This is even more difficult when considering two or more loops. The SZM identity shows that the final result can be obtained in a much simpler  and direct  way.

From the one-loop effective action derived in Reference \cite{Brandt:2012ei} one can easily obtain the pressure of noninteracting thermal particles subjected 
to an external static gravitational field. In a more realistic physical scenario, one would have to take into account
the self-interactions of the thermal particles. In principle this can be investigated by computing the higher-loop contributions, 
which necessarily involves interactions between the thermal particles.
In order to tackle this issue in a systematic fashion, we will investigate in the present work the possibility 
that the SZM identity holds also at the two-loop order for the 1PI diagrams. This issue is also of interest from a broader point of view, 
since there are few higher-loop results at finite temperature, and our analysis is completely general to the extent that
it includes theories with cubic and quartic vertices as well as more general cases, like the weak field expansion of gravity.

It is important to point out that the SZM identity cannot be true when there are external fermionic lines. This is so because it is not possible to make the energy of an external fermionic line equal to zero
before performing the Matsubara sum and the subsequent analytic continuation of the external energy. As a simple explicit example,  
one may consider the static electron self-energy which behaves like $T^2/|\vec k|$ for high-temperatures \cite{lebellac:book96}. Therefore, 
throughout this work, the external lines of the amplitudes are always bosonic. This important limitation shows that the SZM identity is an intrinsic property of the thermal field theories,
since there would be no such distinction between fermionic and bosonic external energies at zero temperature.

In the next section we will review the one-loop calculations which lead to the SZM identity described above. 
We perform an iterative analysis which shows that the one-loop SZM identity 
can be systematically verified for all the $n$-point functions, generalizing the results found in Ref. \cite{Frenkel:2009pi}. 
In Sec. III we consider the two-loop contributions for the two-point function
and explicitly verify the SZM identity for two nontrivial diagrammatic topologies. (There are another three topologies
which are dealt with similarly in the Appendix.) As an example of a trivial topology,
Fig. \ref{fig1} shows a one-loop diagram which does not depend on the external momentum, and therefore satisfies trivially the SZM identity.
(In quantum gravity one should also consider the one-point function which is obviously also independent of the external momentum, at any order in the loop expansion.)
Similarly, the two-loop diagrams in Fig. \ref{fig2} are also independent of the external momentum. 
The topologies shown in Fig. \ref{fig3} can also be considered trivial in the sense that the proof of the SZM identity is similar to the one-loop case.
Finally, in Sec. IV we discuss our results and possible developments.

\begin{figure}[h]
\begin{center}
\includegraphics[scale=0.2]{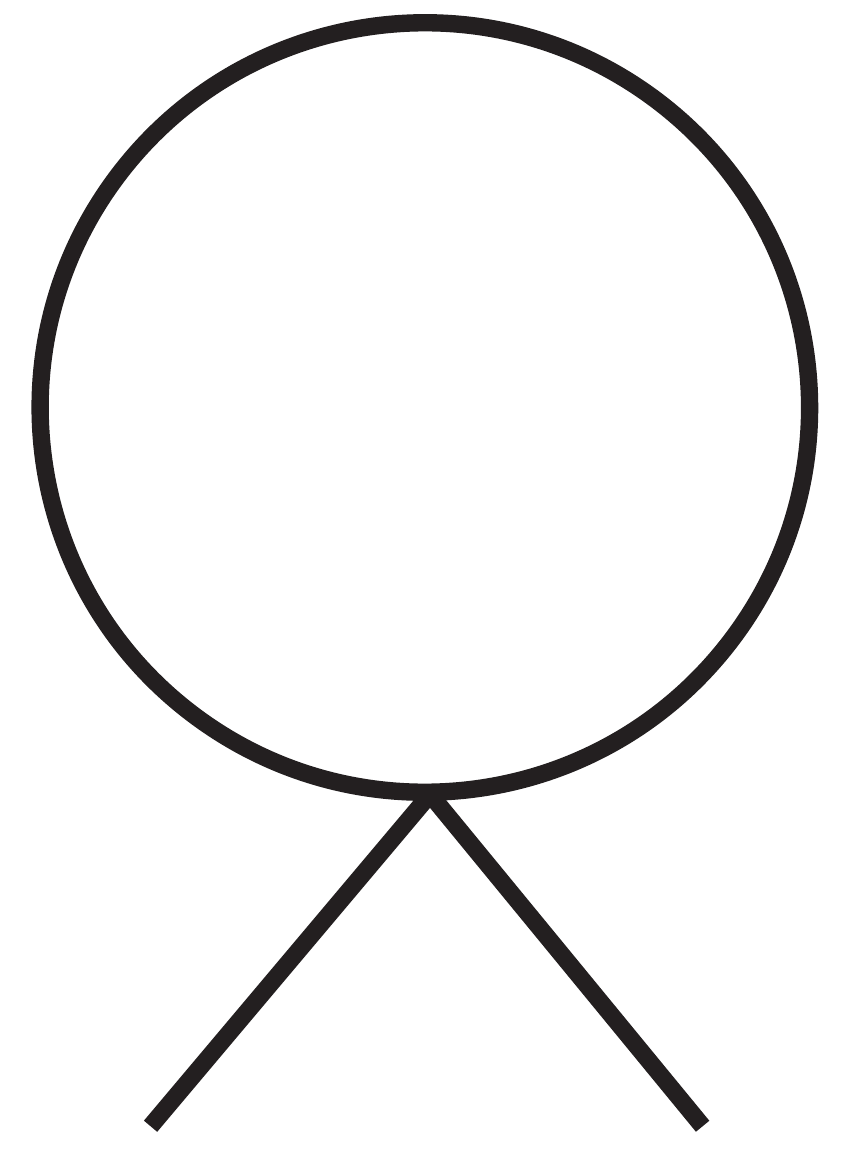}
\caption{Momentum-independent contribution to the two-point function at the one-loop order.}
\label{fig1}
\end{center}
\end{figure}

\begin{figure}[h]
\begin{center}
\includegraphics[scale=0.25]{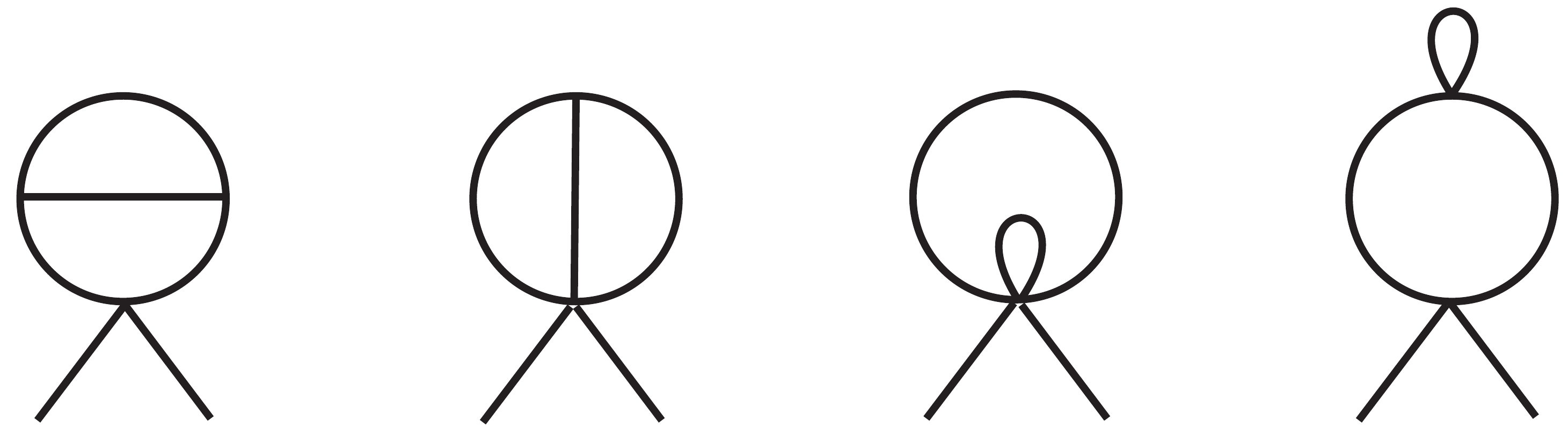}
\end{center}
\caption{Momentum-independent contribution to the two-point function at the two-loop order.}
\label{fig2}
\end{figure}

\begin{figure}[h]
\begin{center}
\includegraphics[scale=0.4]{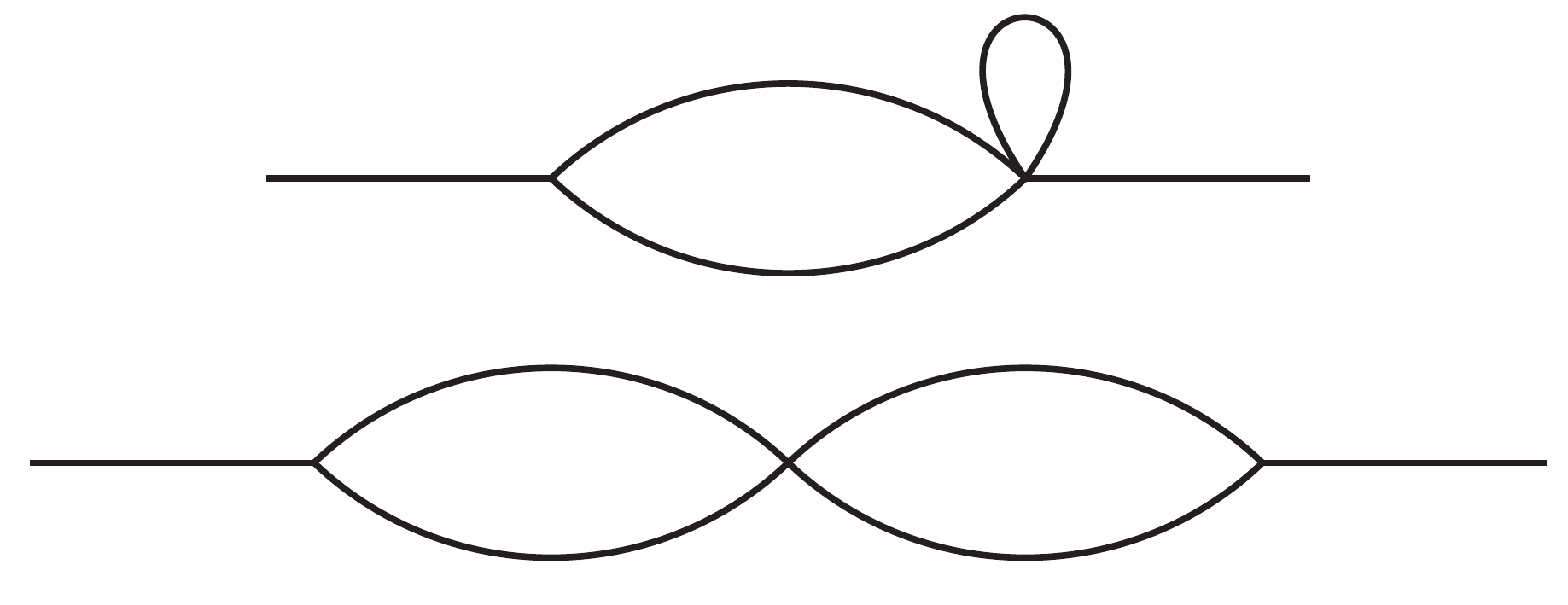}
\end{center}
\caption{Diagrammatic topologies which can be reduced to the one-loop case.}
\label{fig3}
\end{figure}

\section{SZM identity at one-loop order}
\subsection{One-loop self-energy} \label{sec:2}

In a theory of scalar fields, all the diagrammatic topologies depicted in the present work would represent the full 
contribution to a given amplitude. When considering a spinor, vector or tensor theory (like gravity), there may be 
several components associated with a given topology. As will be clear in what follows, for the purpose of our 
present analysis, it is not necessary to make explicit which component we are taking into account. 
All components can be encompassed in the same framework, so that whenever we refer to a given amplitude, we are in 
fact considering several components.

In this subsection we will consider, as a simple example, the one-loop self-energy. This will illustrate our basic idea and
also allow us to introduce the main notation.

\begin{figure}
\begin{center}
\includegraphics[scale=0.4]{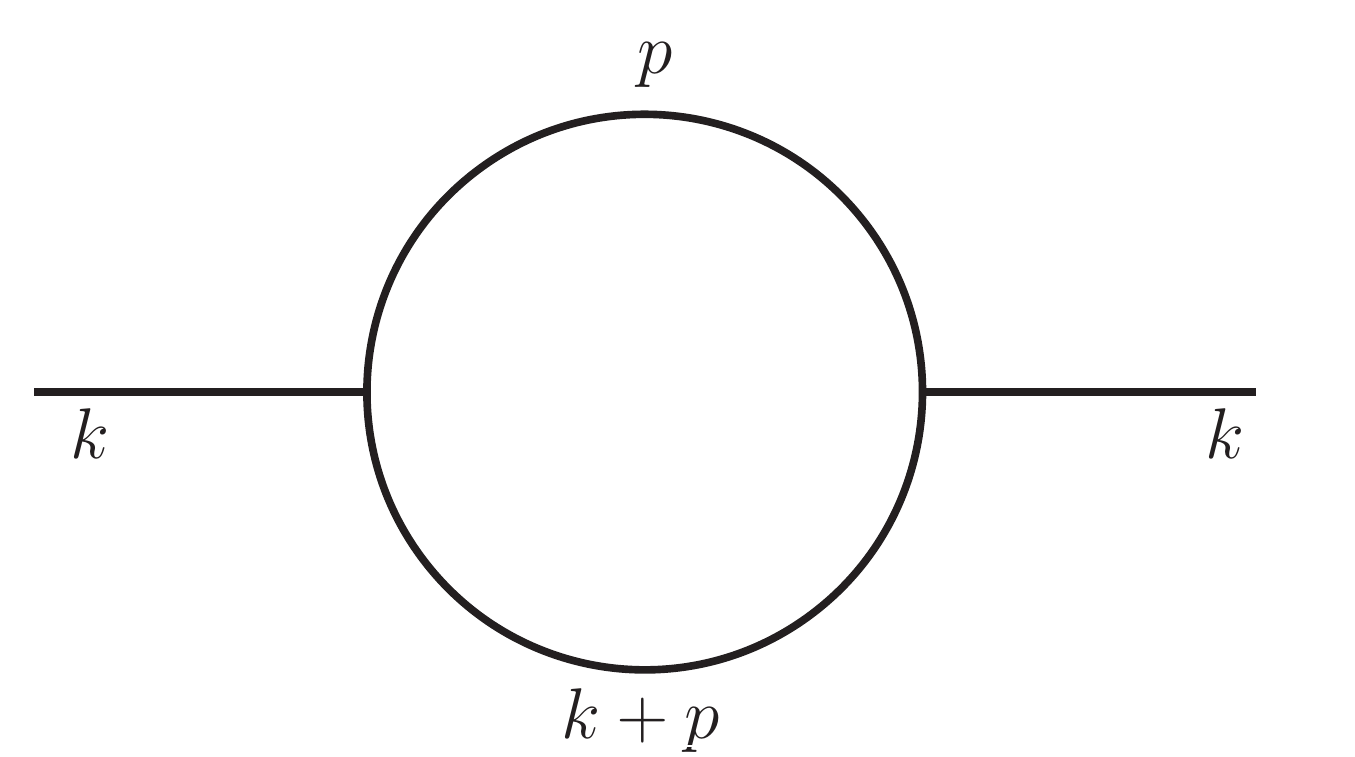}
\caption{A nontrivial one-loop contribution to the self-energy.}
\label{fig:1loop_top1}
\end{center}
\end{figure}

\begin{figure}
\[
\begin{array}{c}
\includegraphics[scale=0.5]{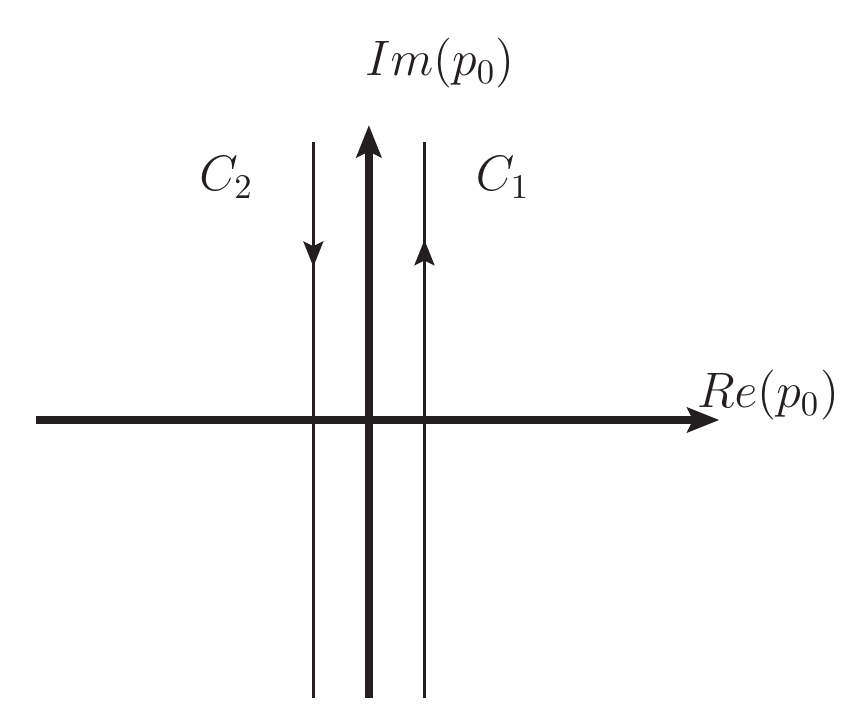} \\ \\ (a)
\\
\includegraphics[scale=0.5]{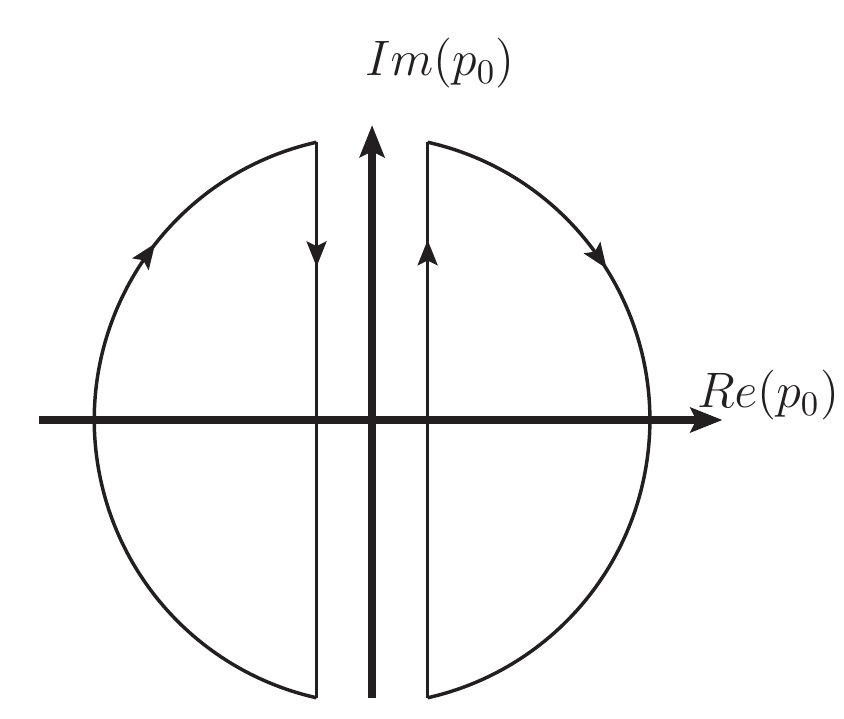} \\ \\ (b)
\end{array}
\]
\caption{Integration paths on the complex plane $p_0$.}
\label{fig:contorno}
\end{figure}

There are two basic topologies which contribute to the self-energy at the one-loop order. The first topology, shown in Fig.~\ref{fig1}, is independent of the external momentum, and the SZM identity is trivial.
The only nontrivial topology which contributes to the one-loop self-energy is shown in Fig.~\ref{fig:1loop_top1}.
In the static limit ($k_0=0$), this amplitude can be written as
\begin{equation}
\Pi_e = \oint_C \frac{dp_0}{2\pi i} \frac{f(p_0)}{(p_0^2-x^2)(p_0^2-w^2)},
\label{eq:pi1loop01}
\end{equation}
where $x=|\vec{p}|$ and $w=|\vec{p}+\vec{k}|$.
(In this work we will omit the integration over the spacial components of the loop momenta since it is not important for the validity of our arguments; 
we remark that this makes our results valid in $d$ space-time dimensions.) 
Here we are using the usual expression for the Matsubara sum in terms of an integral over the contours $C_1$ and $C_2$ shown in Fig. \ref{fig:contorno} (a). 
For the temperature-dependent part, the numerator $f(p_0)$ reduces to
\begin{equation}\label{eq2a}
f(p_0) = N_{p}(p_0)V(p^\mu) .
\end{equation}
Here $N_{p}(p_0)$ is the distribution of Fermi-Dirac or Bose-Einstein, depending on the parity of the line $p_0$.
(In the contour $C_2$ it is convenient to make $f(p_0) \rightarrow -N(-p_0) V(p^\mu)$, which is valid for the temperature-dependent contribution.)

The function $V(p^\mu)$ is model dependent, and it represents some tensor or spinor component. 
In general, the function $V$ depends on the external four-momentum $k$. However, for the Yang-Mills theory or gravity, the components of $V$ are polynomials, and so, in the high-temperature limit, 
we can make $k=0$ in $V$ \cite{lebellac:book96,kapusta:book89}. 
For our present purposes, the only relevant momentum dependence of $f$ is the temporal component of  the momentum $p_0$. 

Using the asymptotic behavior of $N_p(p_0)$, we can close the path $C$ at the infinity as we shown in Fig.~\ref{fig:contorno} (b).
By the residue theorem, we have
\begin{align}
\Pi_e &= - \left[ \frac{f(x)}{2x (x^2 - w^2)} +(x \rightarrow -x) \right. \nonumber \\  &+ \left. \frac{f(w)}{2w(w^2-x^2)} + (w \rightarrow -w) \right] \nonumber \\ &= - \frac{1}{x^2-w^2} \left[ \frac{f(x)}{2x} - \frac{f(w)}{2w}  \right. \nonumber \\ &+ \left. (x,w \rightarrow -x,-w) \right] .
\end{align} 
In the high-temperature limit, the amplitude will be dominated by the region in which the external momentum 
is much smaller than the integration momentum, namely the hard thermal loops (HTL) region \cite{lebellac:book96,kapusta:book89}.  
Therefore, we can write $w = x + \epsilon$, where $\epsilon = w-x \ll x$, and perform an expansion around $\epsilon =0$, so that
\begin{equation}
\frac{1}{x^2-w^2} \left[ \frac{f(x)}{2x} - \frac{f(w)}{2w} \right] \simeq \frac{1}{2x} \frac{d}{d x} \frac{f(x)}{2x}.
\label{eq:aproox00}
\end{equation}
This yields the following contribution to the high-temperature limit of the static self-energy:
\begin{equation}
\Pi_e \simeq -\left[ \frac{1}{2x} \frac{d}{dx} \frac{f(x)}{2x} + (x \rightarrow -x) \right].
\label{eq:frenkelest}
\end{equation}

Let us now consider the other side of the SZM identity, namely the zero-momentum limit. In this case, the self-energy reduces to
\begin{equation}
\Pi_0 =   \oint_C \frac{dp_0}{2\pi i} \frac{f(p_0)}{(p_0^2-x^2)^2}. 
\end{equation}
Using the residue theorem, we can integrate in $p_0$, obtaining
\begin{equation}
\Pi_0 = - \left[ \lim_{p_0 \rightarrow x} \frac{d}{d p_0}  \frac{f(p_0)}{(p_0+x)^2} + (x \rightarrow -x) \right].
\end{equation}
Evaluating the limit, we obtain
\begin{align}
&\lim_{p_0 \rightarrow x} \frac{d}{d p_0}  \frac{f(p_0)}{(p_0+x)^2} = \frac{1}{4 x^2} \frac{d}{dx} f(x) - \frac{f(x)}{4x^3} \nonumber \\  &= \frac{1}{2x} \frac{d}{dx} \frac{f(x)}{2x}.
\end{align}
Therefore, the zero-momentum limit reduces to
\begin{equation}
\Pi_0 = - \left[  \frac{1}{2x} \frac{d}{dx} \frac{f(x)}{2x} + (x \rightarrow -x) \right],
\label{eq:frenkelzero}
\end{equation}
which is identical to the result in Eq. \eqref{eq:frenkelest}. This shows that the contribution of the one-loop nontrivial topology 
in Fig. \ref{fig:1loop_top1} satisfies the SZM identity. Taking into account also the trivial contribution in Fig. \ref{fig1}, 
we conclude that the one-loop self-energy satisfies the SZM identity.

From the explicit expressions for the diagrams shown in Fig.~\ref{fig3} we can see that the 
SZM identity follows from a similar derivation as above. Indeed, in this type of diagram the loop integrals are 
independent of each other.

\subsection{Iterative procedure for one-loop amplitudes}

\begin{figure}[h]
\begin{center}
\includegraphics[scale=0.45]{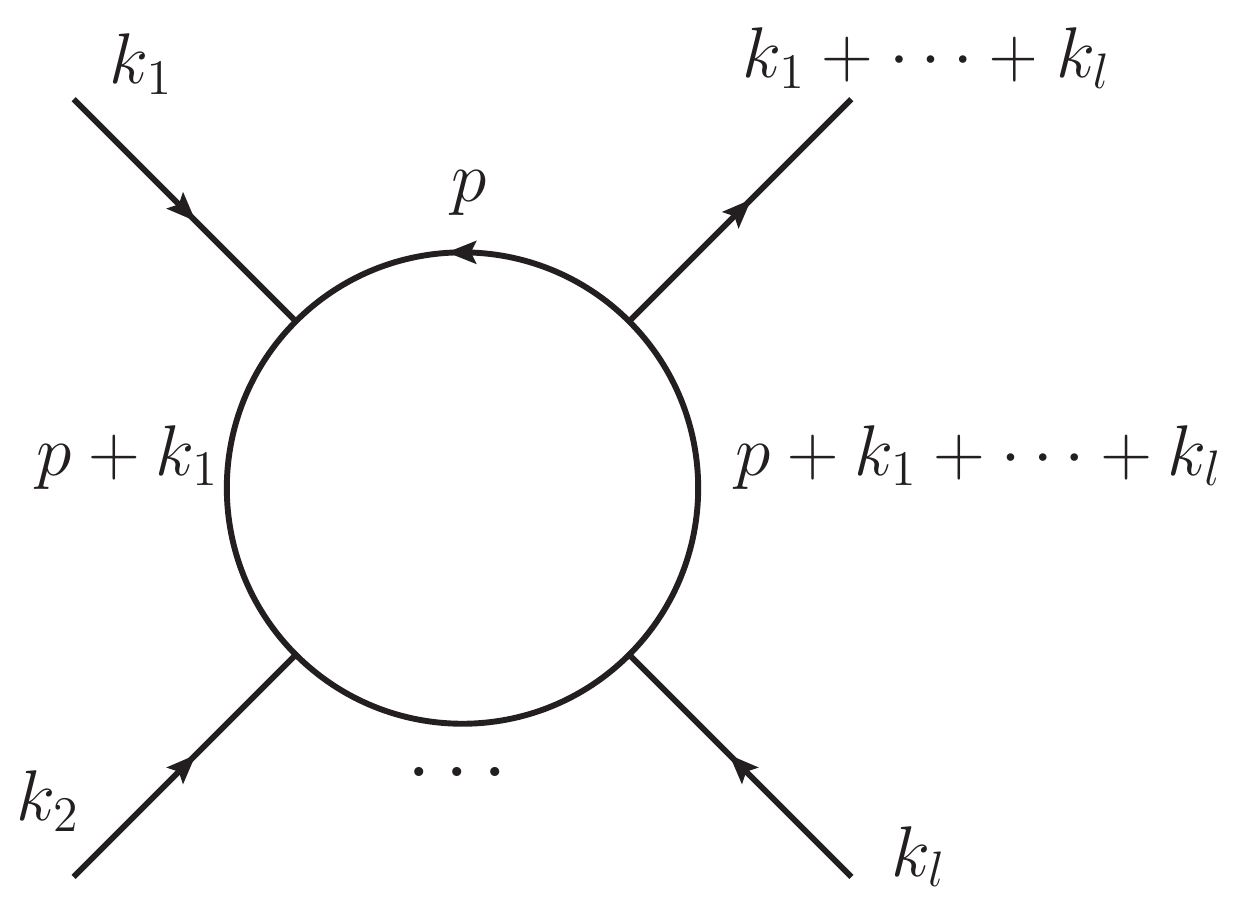}
\caption{1PI one-loop amplitude with $l+1$ vertices.
It is to be understood that some of the external lines may represent a bundle of lines which intersect on the same vertex.}
\label{fig:1pigeral}
\end{center}
\end{figure}

Let us now consider all possible 1PI one-loop amplitudes with an arbitrary number of external lines. 
Fig.  \ref{fig:1pigeral} shows a diagram containing $l+1$ vertices. For a general diagrammatic topology, 
$k_i$ denotes the sum of $n_i = 1,\,2\,\dots$ individual four-momenta associated with the $n_i$ lines 
intersecting on a ($n_i+2$)-point vertex, so that the total number of external lines is given by $\sum_i n_i \ge  l+1$.

In the static limit, when the time component of all external four-momenta vanishes (${k_0}_i =0$), this amplitude can be written as
(recall that we are not concerned with the integrations over the space components of the loop momentum)
\begin{align}
\mathcal{A}^{l+1}_e = \oint_C \frac{dp_0}{2\pi i} \frac{{\bar{f}}(p_0)}{(p_0^2-x^2)}  \prod_{n=1}^{l} \frac{1}{p_0^2 - w_n^2 },
\label{eq:1pigeral}
\end{align} 
where we have introduced the notation
\begin{align}
w_n = \left|\vec{p}+\vec{k}_1 + \cdots + \vec{k}_n\right|,
\end{align}
and $x = \left|\vec{p}\right|$. At high-temperature, the function ${\bar{f}}(p_0)$ has similar properties to $f(p_0)$ 
in Eq. \eqref{eq2a}, being the product of the thermal distribution and  components of a tensor independent of the external momentum.

Performing the $p_0$ integration with the help of the residue theorem, and using the contour depicted in the Fig. \ref{fig:contorno} (b), we obtain
 \begin{align}\label{eq12a}
\mathcal{A}^{l+1}_e &= - \left\{ \frac{{\bar{f}}(x)}{2x} \prod_{n=1}^{l} \frac{1}{x^2 - w_n^2} \right. \nonumber \\&+ \left. \sum_{m=1}^{l} \frac{{\bar{f}}(w_m)}{2w_m (w_m^2-x^2)} \prod_{\nntstile{n\neq m}{n=1}}^{l} \frac{1}{w_m^2-w_n^2}\right\} \nonumber \\
&+\{(x,w_j) \rightarrow (-x,-w_j)\} \;\; j=1\dots l.
\end{align} 
In order to attain the full high-temperature limit of the static amplitude, we now consider an iterative procedure. 
Our strategy, in order to reach the HTL region, is to perform a step-by-step procedure considering each external momentum $\vec k_i$ small compared with $\vec p$. 

Let us first single out the dependence on $w_1$, so that Eq. \eqref{eq12a} can be written as
\begin{widetext}
\begin{align}
\mathcal{A}^{l+1}_e&= -\left\{ \frac{1}{x^2-w_1^2} \left( \frac{{\bar{f}}(x)}{2x} \prod_{n=2}^{l} \frac{1}{x^2 - w_n^2} - \frac{{\bar{f}}(w_1)}{2w_1} \prod_{n=2}^{l} \frac{1}{w_1^2 - w_n^2} \right) +  \sum_{m=2}^{l} \frac{{\bar{f}}(w_m)}{2w_m (w_m^2-x^2)(w_m^2-w_1^2)} \prod_{\nntstile{n\neq m}{n=2}}^{l} \frac{1}{w_m^2-w_n^2}\right\}   \nonumber  \\
&+\{(x,w_j) \rightarrow (-x,-w_j)\} \;\; j=1\dots l
\end{align}
\end{widetext}
We now consider the external momentum $\vec k_1$  much smaller than the loop momentum $\vec{p}$, so that, 
proceeding as in the previous sub-section, we can write $w_1 = x + \epsilon$ and expand around $\epsilon =0$. This yields
\begin{align}
\mathcal{A}^{l+1}_e&\simeq -\left\{ \frac{1}{2x} \frac{\partial}{\partial x} \left( \frac{{\bar{f}}(x)}{2x} \prod_{n=2}^{l} \frac{1}{x^2 - \bar{w}_n^2} \right) \right. \nonumber \\ &+ \left.  \sum_{m=2}^{l} \frac{{\bar{f}}(\bar{w}_m)}{2\bar{w}_m (\bar{w}_m^2-x^2)^2} \prod_{\nntstile{n\neq m}{n=2}}^{l} \frac{1}{\bar{w}_m^2-\bar{w}_n^2}\right\} \nonumber  \\
&+\{(x,\bar{w}_j) \rightarrow (-x,-\bar{w}_j)\} \;\; j=2\dots l,
\label{eq:al+11}
\end{align} 
with the notation
\begin{align}
\bar{w}_n = \left| \vec{p} + \vec{k}_2 + \cdots + \vec{k}_n \right| \;\; n=2 \dots l.
\end{align}
Using the residue theorem we can write Eq. \eqref{eq:al+11} as
\begin{align}
\mathcal{A}^{l+1}_e&\simeq 
\frac{1}{2x} \frac{\partial}{\partial x} \oint_C \frac{dp_0}{2\pi i} \frac{{\bar{f}}(p_0)}{(p_0^2-x^2)}  \prod_{n=2}^{l} \frac{1}{p_0^2 - \bar{w}_n^2 } \nonumber \\
&=\oint_C \frac{dp_0}{2\pi i} \frac{{\bar{f}}(p_0)}{(p_0^2-x^2)^2}  \prod_{n=2}^{l} \frac{1}{p_0^2 - \bar{w}_n^2 }.
\label{eq:basis4induciton}
\end{align}
On the other hand, if we make $\vec k_1=0$ in the static amplitude [Eq. \eqref{eq:1pigeral}], we obtain an amplitude with zero four-momenta ($k_1=0$), which is the same
as Eq. \eqref{eq:basis4induciton}. This special case of the SZM identity is the basis for the inductive reasoning to be employed in what follows. Notice that, in the particular case where $l=1$, 
Eqs. \eqref{eq:1pigeral} (with $\vec k_1=0$) and Eq. \eqref{eq:basis4induciton} explicitly verify the SZM identity for the self-energy, as in the previous section.

In order to complete the inductive reasoning, let us now assume that the amplitude, with $\vec{k}_1, \,\, \dots, \,\, \vec{k}_{n-1}$ much smaller than the loop momentum 
$\vec{p} $, can be approximated by
\begin{align}
\mathcal{A}^{l+1}_e&\simeq \oint_C \frac{dp_0}{2\pi i} \frac{{\bar{f}}(p_0)}{(p_0^2-x^2)^n}  \prod_{i=n}^{l} \frac{1}{p_0^2 - v_i^2 },
\label{eq:indu000}
\end{align}
where
\begin{align}
v_i = \left| \vec{p} + \vec{k}_n + \cdots + \vec{k}_i \right| \;\; i=n,\;\; \dots , \;\; l .
\end{align}
Then, starting from Eq. \eqref{eq:indu000}, we will prove that, when  the momentum $\vec{k}_n$ is also much smaller than $\vec{p}$, we obtain Eq. \eqref{eq:indu000} with 
$n\rightarrow n+1$ [see Eqs. \eqref{eq32a} \eqref{eq36a}].


Using the residue theorem, the  $p_0$ integration in Eq. \eqref{eq:indu000} yields
\begin{widetext}
\begin{align}
\mathcal{A}^{l+1}_e&\simeq -\left\{ \lim_{p_0 \rightarrow x} \frac{1}{(n-1)!} \left(\frac{\partial}{\partial p_0} \right)^{n-1} \frac{\bar{f}(p_0)}{(p_0+x)^n} \prod_{i=n}^l \frac{1}{p_0^2-v_i^2}  + \sum_{j=n}^l \frac{\bar{f}(v_j)}{2v_j (v_j^2-x^2)^n } \prod_{\nntstile{i \neq j}{i=n}}^l \frac{1}{v_j^2-v_i^2}  \right\} \nonumber  \\ 
&+ \{(x,v_m) \rightarrow (-x,-v_m)\} \;\; m=2\dots l, \nonumber \\
&= -\left\{ \lim_{p_0 \rightarrow x} \frac{1}{(n-1)!} \left(\frac{\partial}{\partial p_0} \right)^{n-1} \frac{\bar{f}(p_0)}{(p_0+x)^n (p_0^2 - v_n^2)} \prod_{i=n+1}^l \frac{1}{p_0^2-v_i^2} +  \frac{\bar{f}(v_n)}{2v_n (v_n^2-x^2)^n } \prod_{i=n+1}^l \frac{1}{v_n^2-v_i^2}  \right. \nonumber \\ &+ \left. 
\sum_{j=n+1}^l \frac{\bar{f}(v_j)}{2v_j (v_j^2-x^2)^n(v_j^2 - v_n^2) } \prod_{\nntstile{i \neq j}{i=n+1}}^l \frac{1}{v_j^2-v_i^2} \right\}  + \{(x,v_m) \rightarrow (-x,-v_m)\} \;\; m= n, \;\; \dots, \;\; l. 
\label{eq:induc}
\end{align}
\end{widetext}
The first term in Eq. \eqref{eq:induc} can be written as 
\begin{align}
T_1 &= \lim_{p_0 \rightarrow x} \frac{1}{(n-1)!} \left(\frac{\partial}{\partial p_0} \right)^{n-1} \frac{h(p_0)}{(p_0+x)^n(p_0^2 - v_n^2)}  \nonumber \\
&=  \lim_{p_0 \rightarrow x} \frac{1}{(n-1)!} \left(\frac{\partial}{\partial p_0} \right)^{n-1} \nonumber \\ 
&\times   \frac{h(p_0)}{(p_0+x)^n(p_0+v_n)(p_0 - v_n)},
\label{eq:t1t1}
\end{align}
where we have introduced
\begin{align}\label{eq21a}
h(p_0) = \bar{f}(p_0) \prod_{i=n+1}^l \frac{1}{p_0^2-v_i^2} .
\end{align}
Using the Leibniz rule, we have an analogy with the  binomial formula, such that
\begin{widetext}
\begin{align}
\left(\frac{\partial}{\partial p_0} \right)^{n-1} [A(p_0) B(p_0) ] = \sum_{t=0}^{n-1} \frac{(n-1)!}{t!(n-1-t)!} 
\left[ \left(\frac{\partial}{\partial p_0} \right)^{t} A(p_0) \right] \left[ \left(\frac{\partial}{\partial p_0} \right)^{n-1-t} B(p_0) \right],
\end{align}
so that Eq. \eqref{eq:t1t1} acquires the form
\begin{align}
T_1 &=  \sum_{t=0}^{n-1} \frac{1}{t!(n-1-t)!} \lim_{p_0 \rightarrow x} \left[ \left(\frac{\partial}{\partial p_0} \right)^{n-1-t} \frac{1}{p_0 -v_n}  \right] \left[ \left(\frac{\partial}{\partial p_0} \right)^{t} \frac{h(p_0)}{(p_0+x)^n ( p_0 + v_n)}  \right].
\end{align}
Evaluating the derivative yields
\begin{align}\label{eq24a}
T_1 &=  \sum_{t=0}^{n-1} \frac{1}{t!(n-1-t)!} \lim_{p_0 \rightarrow x} \left[ \frac{(-1)^{n-t-1} (n-t-1)! }{(p_0 -v_n)^{n-t} }  \right] \left[ \left(\frac{\partial}{\partial p_0} \right)^{t} \frac{h(p_0)}{(p_0+x)^n ( p_0 + v_n)}  \right] \nonumber \\
&=  \sum_{t=0}^{n-1} \frac{1}{t!} \lim_{p_0 \rightarrow x} \left[ \frac{-1}{(v_n-p_0)^{n-t} }  \right] \left[ \left(\frac{\partial}{\partial p_0} \right)^{t} \frac{h(p_0)}{(p_0+x)^n ( p_0 + v_n)}  \right] \nonumber \\
&=  \sum_{t=0}^{n-1} \frac{1}{t!} \frac{-1}{(v_n -x)^{n-t} }  \lim_{p_0 \rightarrow x} \left(\frac{\partial}{\partial p_0} \right)^{t} \frac{h(p_0)}{(p_0+x)^n ( p_0 + v_n)}.
\end{align}
\end{widetext}
We now introduce the high-temperature limit, when momentum $\vec{k}_n$ is much smaller than the integration momentum $\vec{p}$, so that $v_n = x +\epsilon$.  Then, the dominant term in Eq. \eqref{eq24a} is
\begin{align}\label{eq25a}
T_1 &\simeq - \sum_{t=0}^{n-1} \frac{1}{t!(v_n -x)^{n-t} }  \lim_{p_0 \rightarrow x} \left(\frac{\partial}{\partial p_0} \right)^{t} \frac{h(p_0)}{(p_0+x)^{n+1}}.
\end{align}

Similarly, using Eq. \eqref{eq21a}, the second term in Eq. \eqref{eq:induc} can be written as
\begin{align}
T_2 = \frac{1}{(v_n-x)^n} \frac{h(v_n)}{2v_n (v_n + x)^n}.
\end{align}
In the high-temperature limit, we can make the approximation
\begin{align}
T_2 \simeq \frac{1}{(v_n-x)^n} \frac{h(v_n)}{(v_n + x)^{n+1}}
\end{align}
and perform the Taylor expansion, yielding
\begin{widetext}
\begin{align}
T_2 \simeq \frac{1}{(v_n-x)^n} \sum_{t=0}^{\infty} \frac{(v_n - x)^t}{t!} \lim_{v_n \rightarrow x} \left( \frac{\partial}{\partial v_n} \right)^t\frac{h(v_n)}{(v_n + x)^{n+1}},
\end{align}
\end{widetext}
which can be rewritten as
\begin{align}\label{eq30a}
T_2 \simeq \sum_{t=0}^{\infty} \frac{1}{t!(v_n - x)^{n-t}} \lim_{p_0 \rightarrow x} \left( \frac{\partial}{\partial p_0} \right)^t\frac{h(p_0)}{(p_0 + x)^{n+1}}.
\end{align}
Combining Eqs. \eqref{eq25a} and \eqref{eq30a} and neglecting all the subleading contributions in the high-temperature limit, we obtain
\begin{align}\label{eq31a}
& T_1+T_2 \simeq \frac{1}{n!} \lim_{p_0 \rightarrow x}  \left( \frac{\partial}{\partial p_0} \right)^n \frac{h(p_0)}{(p_0+x)^{n+1}}  \nonumber \\ 
& \simeq \frac{1}{n!} \lim_{p_0 \rightarrow x}  \left( \frac{\partial}{\partial p_0} \right)^n \frac{\bar{f}(p_0)}{(p_0+x)^{n+1}}\prod_{{i=n+1}}^l \frac{1}{p_0^2-\bar{v}_i^2},
\end{align}
where
\begin{align}\label{eq32a}
\bar{v}_i = \left| \vec{p} + \vec{k}_{n+1} + \cdots + \vec{k}_{i} \right| \,\, , i=n+1 , \,\, \cdots, \,\, l  .
\end{align}

The last term in Eq. \eqref{eq:induc} has the following
high-temperature limit
\begin{align}\label{eq34a}
T_3 \simeq \sum_{j=n+1}^l \frac{\bar{f}(\bar{v}_j)}{2\bar{v}_j (\bar{v}_j^2-x^2)^{n+1} } \prod_{\nntstile{i \neq j}{i=n+1}}^l \frac{1}{\bar{v}_j^2-\bar{v}_i^2}.
\end{align}
Combining Eqs. \eqref{eq31a} and \eqref{eq34a}, we obtain
\begin{widetext}
\begin{align}
\mathcal{A}^{l+1}_e &\simeq - \left\{  \frac{1}{n!} \lim_{p_0 \rightarrow x}  \left( \frac{\partial}{\partial p_0} \right)^n \frac{\bar{f}(p_0)}{(p_0+x)^{n+1}}\prod_{{i=n+1}}^l \frac{1}{p_0^2-\bar{v}_i^2} +\sum_{j=n+1}^l \frac{\bar{f}(\bar{v}_j)}{2\bar{v}_j (\bar{v}_j^2-x^2)^{n+1} } \prod_{\nntstile{i \neq j}{i=n+1}}^l \frac{1}{\bar{v}_j^2-\bar{v}_i^2}  \right\} \nonumber \\
&+ \{(x,\bar{v}_m) \rightarrow (-x,-\bar{v}_m)\} \;\; m= n+1, \;\; \dots, \;\; l.
\end{align}
\end{widetext}
Finally, using the residue theorem yields
\begin{align}\label{eq36a}
\mathcal{A}^{l+1}_e &\simeq \oint_C \frac{dp_0}{2\pi i} \frac{{\bar{f}}(p_0)}{(p_0^2-x^2)^{n+1}}  \prod_{i=n+1}^{l} \frac{1}{p_0^2 - \bar{v}_i^2 }.
\end{align}
This completes the inductive reasoning for all one-loop HTL amplitudes.

As a direct consequence of the previous proof, we now can write
\begin{align}
\mathcal{A}^{l+1}_e \simeq \oint_C \frac{dp_0}{2 \pi i} \frac{ \bar{f}(p_0)}{(p_0^2 - x^2)^{l+1}} = \mathcal{A}^{l+1}_0,
\end{align}
which generalizes the result obtained in Ref. \cite{Frenkel:2009pi}. Then, from this general SZM identity, we can immediately write
the result for a general 1PI one-loop static amplitude, in the HTL approximation, as follows:
\begin{align}
\mathcal{A}^{l+1}_e &\simeq  \frac{1}{l!}  \left( \frac{1}{2x}\frac{\partial}{\partial x}\right)^l\oint_C \frac{dp_0}{2 \pi i} \frac{ \bar{f}(p_0)}{p_0^2 - x^2} \nonumber \\ 
 &=  \frac{1}{l!}  \left( \frac{1}{2x}\frac{\partial}{\partial x}\right)^l \frac{\bar{f} (x)}{2x} + ( x \rightarrow -x).
\end{align}
This equation gives the general expression for 1PI amplitudes in the high-temperature static limit, after we explicitly
consider some field theory model and integrate over the space components of the loop momentum. It is easy to verify that in $d$ space-time dimensions, 
the power of the temperature will be $[V]+d-2 l -2$ (where $[V]$ denotes the mass dimension of the tensor structure). In the case of scalar ($[V]=0$)  and
vector gauge theories the leading high-temperature behavior will be $T^{d-4}$ and $T^{d-2}$, respectively, while for gravity,  
the 1PI amplitudes contribute to all orders being proportional to $T^d$.

\section{Self-energy at two-loop order}

\begin{figure}[h]
\includegraphics[scale=0.45]{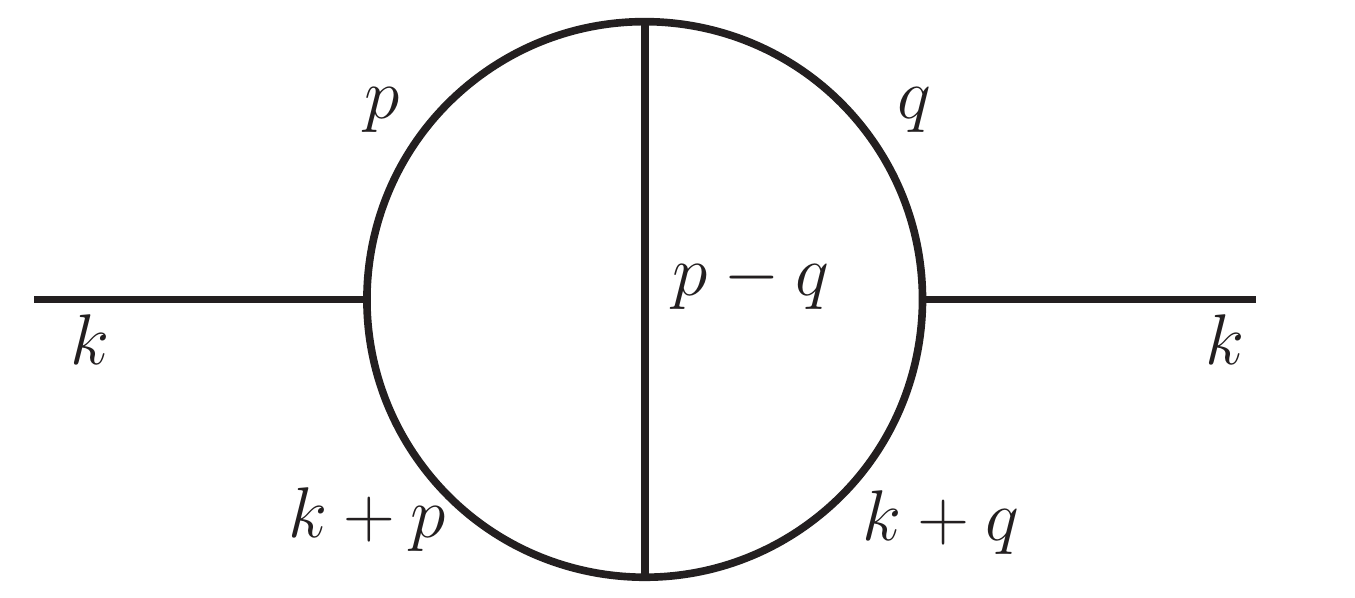}
\caption{First contribution for two-loop self energy}
\label{fig:Toplogia1}
\end{figure}

In this section we will consider the two-loop contributions to the self-energy. 
Now we have to consider all types of nontrivial topologies which, unlike the ones shown in Figs. 
\ref{fig2} and \ref{fig3}, are dependent on the external momentum and cannot be reduced to the one-loop case.
There are five such topologies. Here we will present the details of two of them, which are representative
of the technicalities and illustrate the main aspects of this analysis.
The other three topologies will be considered in the Appendix.

\subsection{First topology}

Let us first consider the contribution shown in Fig.~\ref{fig:Toplogia1}. In the static limit, this amplitude can be written as
\begin{widetext}
\begin{equation}
\Pi_e^A = \oint_C \frac {dp_0}{2\pi i} \oint_C \frac {dq_0}{2\pi i}  \frac{f^A(p_0,q_0)}{ (p_0^2-x^2)(p_0^2-w^2)[(p_0-q_0)^2-z^2](q_0^2-y^2)(q_0^2-v^2)}, 
\label{eq:i0top1}
\end{equation}
where $x=|\vec{p}|$,  $y=|\vec{q}|$, $w=|\vec{p}+\vec{k}|$, $v=|\vec{q}+\vec{k}|$ and $z=|\vec{p}-\vec{q}|$.  
The hard thermal loop region, which yields leading high-temperature contributions, comes from the terms containing the product of two thermal 
distributions so that the function $f^A(p_0,q_0)$ has the following form (there are indications that these terms may be the only ones which survive in the high-temperature limit \cite{Mottola:2009mi})
\begin{align}\label{eqfa}
f^A(p_0,q_0)= N_p(p_0) N_q(q_0) V(p^\mu,q^\nu),
\end{align}
and the same high-temperature considerations we have made for $f(p_0)$ in Sec. ~\ref{sec:2} also holds  for $f^A(p_0,q_0)$, so that it does not depend on the external momentum.

Using the residue theorem to perform the $p_0$ integration yields
\begin{align}
\Pi_e^A  &= -  \oint_C \frac{dq_0}{2\pi i} \left\{ \frac{f^A(x,q_0)} {2x(x^2-w^2)[(x-q_0)^2 -z^2](q_0^2 -y^2)(q_0^2-v^2)} \right. \nonumber \\ &+ \left.  \frac{f^A(w,q_0)} {2w (w^2-x^2)[(w-q_0)^2 -z^2](q_0^2 -y^2)(q_0^2-v^2)}  \right. 
\nonumber \\ &+  \left.  \frac{f^A(q_0+z,q_0)} {2z [(q_0+z)^2-x^2] [(q_0+z)^2-w^2](q_0^2 -y^2)(q_0^2-v^2)} \right\} + \{(x,w,z) \rightarrow (-x,-w,-z) \} .
\end{align}
Similarly, the $q_0$ integral produces
\begin{align}\label{eq25}
\Pi_e^A &=  \left\{\left[\frac{f^A(x,x+z) }{4zx(x^2-w^2)[(x+z)^2-y^2][(x+z)^2-v^2]} + \frac{f^A(w,w+z) }{4zw(w^2-x^2)[(w+z)^2-y^2][(w+z)^2-v^2]} \right.\right. \nonumber \\  &+ \left. \left. (z\rightarrow -z)\right] 
+ \left[ \frac{f^A(x,y)}{4xy(x^2-w^2)[(x-y)^2-z^2](y^2-v^2)} +\frac{f^A(w,y)}{4wy(w^2-x^2)[(w-y)^2-z^2](y^2-v^2)} \right.\right.  \nonumber \\ &+ \left.\left. \frac{f^A(y+z,y)}{4zy[(y+z)^2-x^2][(y+z)^2-w^2](y^2-v^2)}+(y\rightarrow -y) \right] \right. \nonumber \\ 
&+  \left. \left[ \frac{f^A(x,v)}{4xv(x^2-w^2)[(x-v)^2-z^2](v^2-y^2)}  + \frac{f^A(w,v)}{4wv(w^2-x^2)[(w-v)^2-z^2](v^2-y^2)} \right. \right. \nonumber \\ &+ \left. \left.  \frac{f^A(v+z,v)}{4zv[(v+z)^2-x^2][(v+z)^2-w^2](v^2-y^2)} +(v\rightarrow -v) \right]  +  \left[ \frac{f^A(x,x-z)}{4zx(x^2-w^2)[(x-z)^2-y^2][(x-z)^2-v^2]} \right. \right. \nonumber \\  &+ \left. \left. (x\rightarrow -x)\right]  + \left[ \frac{f^A(w,w-z)}{4zw(w^2-x^2)[(w-z)^2-y^2][(w-z)^2-v^2]}+(w\rightarrow -w) \right] \right\} \nonumber \\ &+ \left\{ [(x,w,z)\rightarrow (-x,-w,-z)] \right\} .
\end{align}
\end{widetext}

The HTL region is now characterized by  $q\gg k$ and $p\gg k$, so that we can write
$w=x+\epsilon$ and $v=y+\delta$, with $\epsilon=w-x\ll x$ and $\delta =v-y \ll y$, 
and a series expansion can be performed around $\epsilon=0$ and $\delta=0$.
In this way, all the terms in Eq. \eqref{eq25} can be dealt with using the following relations:
\begin{align}
&\frac{1}{x^2-w^2}\left[\frac{f^A(x,x+z)}{4zx[(x+z)^2-y^2] [(x+z)^2-v^2]} \right. \nonumber \\ & - \left. \frac{f^A(w,w+z)}{4zw[(w+z)^2-y^2] [(w+z)^2-v^2]}  \right] \simeq \nonumber \\ & \simeq \frac{1}{2x} \frac{\partial}{\partial x} \left[\frac{f^A(x,x+z)}{4zx[(x+z)^2-y^2]^2} \right],
\end{align}
\begin{align}
&\frac{1}{y^2-v^2} \left[ \frac{f^A(y+z,y)}{4zy[(y+z)^2-x^2][(y+z)^2- w^2]}\right. \nonumber \\ & - \left.  \frac{f^A(v+z,v)}{4zv[(v+z)^2-x^2][(v+z)^2- w^2]} \right] \nonumber \\
& \simeq \frac{1}{2y}  \frac{\partial}{\partial y} \left[ \frac{f^A(y+z,y)}{4zy[(y+z)^2-x^2]^2} \right] ,
\end{align}
\begin{align}
& \frac{1}{x^2-w^2} \left[ \frac{f^A(x,x-z)}{4zx[(x-z)^2-y^2][(x-z)^2-v^2]} \right. \nonumber \\ & - \left. \frac{f^A(w,w-z)}{4zw[(w-z)^2-y^2][(w-z)^2-v^2]} \right]  \nonumber \\
& \simeq \frac{1}{2x} \frac{\partial}{\partial x}\left[ \frac{f^A(x,x-z)}{4zx[(x-z)^2-y^2]^2} \right] ,
\end{align}
\begin{align}
& \frac{1}{x^2-w^2} \frac{1}{y^2-v^2}  \left[ g(x,y)-g(w,y)-g(x,v)+g(w,v) \right]  \nonumber \\
& \simeq \frac{1}{2x} \frac{\partial}{\partial x}  \frac{1}{y^2-v^2} \left[ g(x,y)-g(x,v)\right] \nonumber \\
& \simeq \frac{1}{4xy} \frac{\partial^2}{\partial x \partial y} g(x,y) ,  
\end{align}
with
\begin{equation}
g(x,y)= \frac{f^A(x,y)}{4xy[(x-y)^2-z^2]}. 
\end{equation}
Substituting these results back into Eq. \eqref{eq25}, we obtain the following static hard thermal loop approximation:
\begin{align}
\Pi_e^A &\simeq \left\{ \frac{1}{2x} \frac{\partial}{\partial x} \left[\frac{f^A(x,x+z)}{4zx[(x+z)^2-y^2]^2} \right] + [z \rightarrow -z] \right. \nonumber \\ &+ \left.   \frac{1}{2x} \frac{\partial}{\partial x}\left[ \frac{f^A(x,x-z)}{4zx[(x-z)^2-y^2]^2} \right] +[x\rightarrow -x] \right. \nonumber \\  &+ \left.  \frac{1}{2y}  \frac{\partial}{\partial y} \left[ \frac{f^A(y+z,y)}{4zy[(y+z)^2-x^2]^2} \right] +[y\rightarrow -y] \right. \nonumber \\ &+ \left. 
\frac{1}{4xy} \frac{\partial^2}{\partial x \partial y}   \frac{f^A(x,y)}{4xy[(x-y)^2-z^2]} + [y \rightarrow -y] \right\} \nonumber \\ &+  \{(x,z) \rightarrow (-x,-z) \}. 
\label{eq:pie1final4}
\end{align}

Let us now consider the zero-momentum limit. In this case Eq. \eqref{eq:i0top1} reduces to
\begin{widetext}
\begin{equation}
\Pi_0^A = \oint_C \frac{dp_0}{2\pi i} \oint_C \frac{dq_0}{2\pi i} \frac{f^A(p_0, q_0)}{(p_0^2-x^2)^2[(p_0-q_0)^2-z^2] (q_0^2-y^2)^2 }.
\end{equation}
Using the residue theorem, the $p_0$ integration yields
\begin{align}
\Pi_0^A &= -\oint_C \frac{dq_0}{2\pi i}  \left\{ \frac{f^A(q_0+z,q_0)}{2z [(q_0+z)^2-x^2]^2 (q_0^2-y^2)^2} + (z \rightarrow -z) + \lim_{p_0 \rightarrow x} \frac{\partial}{\partial p_0} \frac{f^A(p_0,q_0)}{(p_0+x)^2 [(p_0-q_0)^2-z^2](q_0^2-y^2)^2}  \right. \nonumber \\ &+ \left. (x\rightarrow -x) \right\}  \nonumber \\ 
&= - \oint_C \frac{dq_0}{2\pi i} \left\{\left[\frac{f^A(q_0+z,q_0)}{2z [(q_0+z)^2-x^2]^2 (q_0^2-y^2)^2} + \frac{1}{2x} \frac{\partial}{\partial x} \frac{f^A(x,q_0)}{2x [(x-q_0)^2-z^2](q_0^2-y^2)^2} \right] \right. \nonumber \\ &+ \left. [(x,z)\rightarrow(-x,-z)] \right\}.
\end{align}
Similarly, the $q_0$ integral produces 
\begin{align}
\Pi_0^A &= \left\{\lim_{q_0 \rightarrow x-z} \frac{\partial}{\partial q_0} \frac{f^A(q_0+z,q_0)}{2z(q_0+z+x)^2 (q_0^2-y^2)^2} + (x \rightarrow -x) +\lim_{q_0 \rightarrow y } \frac{\partial}{\partial q_0} \frac{f^A(q_0+z,q_0)}{2z[(q_0+z)^2-x^2]^2 (q_0+y)^2} + (y \rightarrow -y)  \right. \nonumber \\ &+ \left. \frac{1}{2x} \frac{\partial}{\partial x } \lim_{q_0\rightarrow y} \frac{\partial}{\partial q_0} \frac{f^A(x,q_0)}{2x [(x-q_0)^2-z^2] (q_0+y)^2 } + (y\rightarrow -y) + \frac{1}{2x} \frac{\partial}{\partial x } \frac{f^A(x,x+z)}{4xz[(x+z)^2-y^2]^2 } +(z\rightarrow -z) \right\}  \nonumber \\ &+ \{(x,z)\rightarrow(-x,-z)\}
\nonumber \\ 
&=  \left\{ \frac{1}{2x} \frac{\partial}{\partial x} \frac{f^A(x,x-z)}{4zx[(x-z)^2-y^2]^2} +(x \rightarrow -x) + \frac{1}{2y}\frac{\partial}{\partial y} \frac{ f^A(y+z,y)}{4zy[(y+z)^2-x^2]^2} + (y\rightarrow -y) \right. \nonumber \\
&+ \left. \frac{1}{4xy}\frac{\partial^2}{\partial x \partial y } \frac{ f^A(x,y) }{4xy [(x-y)^2-z^2] } + (y\rightarrow -y) + \frac{1}{2x} \frac{\partial}{\partial x } \frac{f^A(x,x+z)}{4xz[(x+z)^2-y^2]^2 } +(z\rightarrow-z) \right\} \nonumber \\ &+ \{(x,z)\rightarrow(-x,-z)\},
\label{eq:pi10final4}
\end{align}
which is the same as the static limit [\eqref{eq:pie1final4}].  
This concludes the verification of the SZM identity for the amplitude shown in Fig.~\ref{fig:Toplogia1}.
It is remarkable that this identity holds even before performing the integrations over $\vec p$ and $\vec q$.  

\subsection{Second topology}

The topology shown in Fig. \ref{fig:topologia2} has the following form in the static limit:
\begin{figure}
\begin{center}
\includegraphics[scale=0.45]{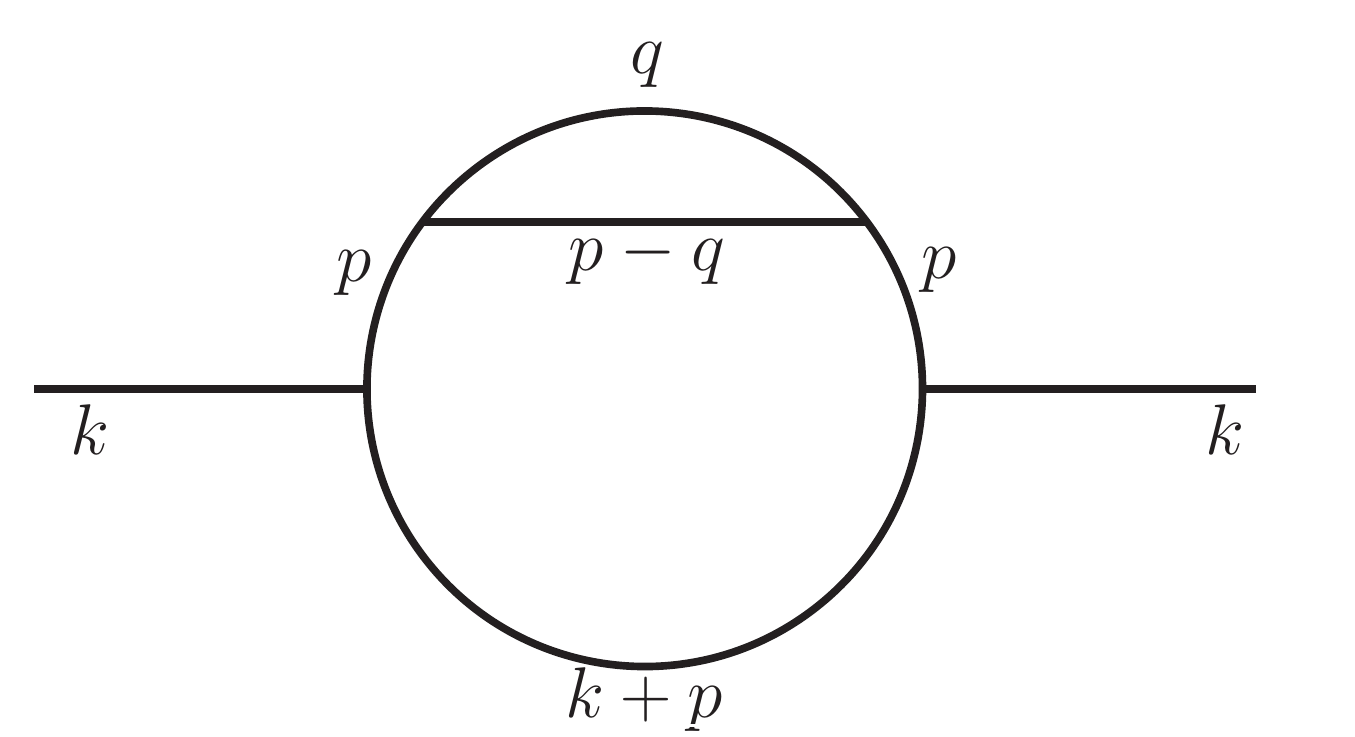}
\end{center}
\caption{Second topology for two-loop self-energy}
\label{fig:topologia2}
\end{figure}
\begin{equation}
\Pi_e^B = \oint_C \frac{dp_0}{2 \pi i} \oint_C \frac{dq_0}{2\pi i} \frac{f^{B}(p_0,q_0)}{ (p_0^2-x^2)^2 (p_0^2 - w^2) [(p_0-q_0)^2-z^2] (q_0^2- y^2)}.
\end{equation}
In this case we introduce the quantities $x=\left|\vec{p}\right|$, $y=\left|\vec{q}\right|$, $w=|\vec{p}+\vec{k}|$ and $z=\left| \vec{p}-\vec{q} \right|$.  
In the high-temperature limit, the numerator $f^B$ has the same form as $f^A$ in Eq. \eqref{eqfa}. 
The integrations over $p_0$ and $q_0$ can be evaluated, as in the previous case, using the residue theorem, yielding
\begin{align}\label{eq40}
\Pi_e^B &= \left\{ \left[ \frac{1}{2x} \frac{\partial}{\partial x} \frac{f^{B}(x,x-z)}{4xz ( x^2 -w^2) [(x-z)^2-y^2]} + (x \rightarrow -x) \right] + \left[ \frac{f^{B}(w,w+z)}{4zw(w^2-x^2)^2[(w+z)^2-y^2]}\right. \right.  \nonumber \\ &+ \left. \frac{1}{2x} \frac{\partial}{\partial x} \frac{f^{B}(x,x+z)}{ 4xz (x^2 - w^2) [(x+z)^2- y^2]} + (z \rightarrow -z) \right] + \left[ \frac{f^{B}(w,w-z)}{4zw(w^2-x^2)^2[(w-z)^2-y^2]} + (w \rightarrow -w) \right]  \nonumber \\ &+ \left[ \frac{f^{B}(y+z,y)}{4zy [(y+z)^2- x^2]^2 [ (y+z)^2 -w^2]} + \frac{f^{B}(w,y)}{4wy (w^2-x^2)^2[(w-y)^2-z^2]} \right. \nonumber \\ &+ \left. \left. \frac{1}{2x} \frac{\partial}{\partial x} \frac{f^{B}(x,y)}{ 4yx (x^2 - w^2) [(x-y)^2-z^2] } +(y \rightarrow -y) \right] \right\} +\{(x,w,z) \rightarrow (-x,-w,-z) \},
\end{align}
where the derivative terms come from doubles poles. In the HTL region we set $w=x+\epsilon$, with $\epsilon \ll x$. Using relations like
\begin{align}
& \frac{g(w)}{(w^2-x^2)^2} +  \frac{1}{2x} \frac{\partial}{\partial x} \frac{ g(x)}{x^2-w^2} \simeq   \frac{g(x)}{(w^2-x^2)^2} + \frac{1}{(w+x)(w^2-x^2)} \frac{\partial}{\partial x} g(x) \nonumber \\ &+ \frac{1}{2(w+x)^2} \frac{\partial^2}{\partial x^2} g(x) - \frac{1}{(x^2-w^2)^2} g(x)  + \frac{1}{2x (x^2-w^2)} \frac{\partial}{\partial x} g(x) \nonumber \\
&\simeq -\frac{1}{8x^3} \frac{\partial}{\partial x} g(x) + \frac{1}{8x^2} \frac{\partial^2}{\partial x^2} g(x),
\label{eq:htldob}
\end{align}
we obtain from \eqref{eq40}
\begin{align}
\Pi_e^B &\simeq \left\{ \left[ \left( \frac{1}{8x^2} \frac{\partial^2}{\partial x^2} -\frac{1}{8x^3} \frac{\partial}{\partial x} \right)  \frac{f^{B}(x,x-z)}{4xz [(x-z)^2-y^2]} + (x \rightarrow -x) \right] + \left[ \left( \frac{1}{8x^2} \frac{\partial^2}{\partial x^2} -\frac{1}{8x^3} \frac{\partial}{\partial x} \right) \frac{f^{B}(x,x+z)}{ 4xz  [(x+z)^2- y^2]} \right. \right. \nonumber \\  &+ \left. \left. (z \rightarrow -z) \right]  + \left[  \left( \frac{1}{8x^2} \frac{\partial^2}{\partial x^2} -\frac{1}{8x^3} \frac{\partial}{\partial x} \right) \frac{f^{B}(x,y)}{ 4yx [(x-y)^2-z^2] } +\frac{f^{B}(y+z,y)}{4zy [(y+z)^2- x^2]^3 }+(y \rightarrow -y) \right] \right\} \nonumber \\ 
&+ \{(x,z) \rightarrow (-x,-z)\}.
\end{align}
It is also convenient to employ the identity
\begin{align}
\left( \frac{1}{8x^2} \frac{\partial^2}{\partial x^2} -\frac{1}{8x^3} \frac{\partial}{\partial x} \right) \frac{h(x)}{x} = 
\left( \frac{3}{8x^5}-\frac{3}{8x^4}\frac{\partial}{\partial x} +\frac{1}{8x^3} \frac{\partial^2}{\partial x^2} \right) h(x),
\label{eq:iiddjjbb}
\end{align}
so that the amplitude takes the final form  
\begin{align}
\Pi_e^B &\simeq \left\{ \left[\left( \frac{3}{8x^5}-\frac{3}{8x^4}\frac{\partial}{\partial x} +\frac{1}{8x^3} \frac{\partial^2}{\partial x^2} \right) \frac{f^{B}(x,x-z)}{4z [(x-z)^2-y^2]} + (x \rightarrow -x) \right] \right. \nonumber \\ &+ \left. \left[\left( \frac{3}{8x^5}-\frac{3}{8x^4}\frac{\partial}{\partial x} +\frac{1}{8x^3} \frac{\partial^2}{\partial x^2} \right) \frac{f^{B}(x,x+z)}{ 4z  [(x+z)^2- y^2]}+ (z \rightarrow -z) \right] \right. \nonumber \\  &+ \left. \left[  \left( \frac{3}{8x^5}-\frac{3}{8x^4}\frac{\partial}{\partial x} +\frac{1}{8x^3} \frac{\partial^2}{\partial x^2} \right) \frac{f^{B}(x,y)}{ 4y [(x-y)^2-z^2] } +\frac{f^{B}(y+z,y)}{4zy [(y+z)^2- x^2]^3 }+(y \rightarrow -y) \right] \right\} \nonumber \\ 
&+ \{(x,z) \rightarrow (-x,-z)\}.
\label{eq:pie2ff}
\end{align}

Let us now consider the zero-momentum limit. In this case, the amplitude shown in Fig.~\ref{fig:topologia2} reduces to
\begin{equation}
\Pi_0^B = \oint_C \frac{dp_0}{2 \pi i} \oint_C \frac{dq_0}{2\pi i} \frac{f^{B}(p_0,q_0)}{ (p_0^2-x^2)^3[(p_0-q_0)^2-z^2] (q_0^2- y^2)}.
\end{equation}
Integrating in $p_0$,
\begin{align}
\Pi_0^B &=- \oint_C \frac{dq_0}{2 \pi i}  \left\{\frac{f^{B}(q_0+z,q_0)}{2z [(q_0+z)^2- x^2]^3 (q_0^2-y^2)} +(z \rightarrow -z)\right. \nonumber \\ &+ \left. \frac{1}{2}\lim_{p_0\rightarrow x} \frac{\partial^2}{\partial p_0^2} \frac{f^{B}(p_0,q_0)}{(p_0+x)^3 [(p_0-q_0)^2-z^2] (q_0^2-y^2)} + (x \rightarrow -x) \right\} 
\end{align}
and computing the limit, yields 
\begin{align}
\Pi_0^B &= \oint_C \frac{dq_0}{2 \pi i}  \left\{\left[\frac{f^{B}(q_0+z,q_0)}{2z [(q_0+z)^2- x^2]^3 (q_0^2-y^2)} +  \left( \frac{3}{16x^5}-\frac{3}{16x^4}\frac{\partial}{\partial x} +\frac{1}{16x^3} \frac{\partial^2}{\partial x^2} \right)\frac{f^{B}(x,q_0)}{ [(x-q_0)^2-z^2] (q_0^2-y^2)}\right] \right. \nonumber \\  
&+ \{ (x,z) \rightarrow (-x,-z) \} .
\end{align}
Finally, integrating in $q_0$, we obtain
\begin{align}
\Pi_0^B &= \left\{ \left[ \frac{1}{2} \lim_{q_0 \rightarrow x-z} \frac{\partial^2}{\partial q_0^2}  \frac{f^{B}(q_0+z,q_0)}{2z (q_0+z+ x)^3 (q_0^2-y^2)} +( x \rightarrow -x) \right] + \left[  \frac{f^{B}(y+z,y)}{4zy [(y+z)^2- x^2]^3} + ( y \rightarrow -y) \right] \right. \nonumber \\
&+ \left[  \left( \frac{3}{16x^5}-\frac{3}{16x^4}\frac{\partial}{\partial x} +\frac{1}{16x^3} \frac{\partial^2}{\partial x^2} \right)\frac{f^{B}(x,y)}{ 2y [(x-y)^2-z^2] } + (y \rightarrow -y ) \right] \nonumber \\ &+ \left. \left[  \left( \frac{3}{16x^5}-\frac{3}{16x^4}\frac{\partial}{\partial x} +\frac{1}{16x^3} \frac{\partial^2}{\partial x^2} \right)\frac{f^{B}(x,x+z)}{2z [(x+z)^2-y^2]} + (z\rightarrow -z) \right] \right\} +\{(x,z) \rightarrow (-x,-z) \}  \nonumber \\
 &= \left\{ \left[\left( \frac{3}{16x^5}-\frac{3}{16x^4}\frac{\partial}{\partial x} +\frac{1}{16x^3} \frac{\partial^2}{\partial x^2} \right)  \frac{f^{B}(x,x-z)}{2z  [(x-z)^2-y^2]} +( x \rightarrow -x) \right] + \left[  \frac{f^{B}(y+z,y)}{4zy [(y+z)^2- x^2]^3} + ( y \rightarrow -y) \right] \right. \nonumber \\
&+ \left[  \left( \frac{3}{16x^5}-\frac{3}{16x^4}\frac{\partial}{\partial x} +\frac{1}{16x^3} \frac{\partial^2}{\partial x^2} \right)\frac{f^{B}(x,y)}{ 2y [(x-y)^2-z^2] } + (y \rightarrow -y ) \right] \nonumber \\ &+ \left. \left[  \left( \frac{3}{16x^5}-\frac{3}{16x^4}\frac{\partial}{\partial x} +\frac{1}{16x^3} \frac{\partial^2}{\partial x^2} \right)\frac{f^{B}(x,x+z)}{2z [(x+z)^2-y^2]} + (z\rightarrow -z) \right] \right\} +\{(x,z) \rightarrow (-x,-z) \},
\label{eq:pi02ff}
\end{align}
which is identical to the static-limit result in Eq.~\eqref{eq:pie2ff}, in agreement with the SZM identity.

\end{widetext}
\bigskip

\section{Discussion}

The results presented in this work may be useful in physical scenarios where there are static external fields
and the temperature is high. An important example would arise in the calculation of
effective actions in static backgrounds. Indeed, in the configuration space, the SZM identity implies that
one may compute the static effective action using a much simpler space-time independent background field configuration.
For instance in Ref. \cite{Brandt:2012ei} we have shown that the effective action of static gravitational 
fields can be obtained, in a closed-form, using the hypothesis that the external fields are space-time independent. 
We based this hypothesis on the results suggested in \cite{Frenkel:2009pi}, 
where the SZM identities was verified up to the three-point function. 
In the present work, we have provided an iterative general proof of the one-loop SZM for all $n$-point 1PI one-loop Green's functions. 

The main result of the present paper is the proof of the SZM identity, using a rather lengthy calculation, at the two-loop order, in the
case of the two-point function. The two-loop result is physically interesting because, at finite temperature, it may be employed in order to describe 
physical situations such that the thermal particles are interacting not only with the external fields, but also with each other.

It is important to point out that we have proved the SZM identity for the hard thermal loop contributions to the thermal amplitudes. 
Nonleading contributions to the static amplitudes would in general depend on the scale of external three-momenta or the mass (for massive field theories). 
While it is obvious that any subleading contribution which depends on the external momenta would violate the SZM identity,
the same may not be true for the mass-dependent subleading terms. However, this would only be relevant (if true) for the nonleading contributions to static amplitudes.

We also remark that the Feynman graph topologies which we have considered in this work are sufficiently general to encompass 
a rather general class of field theories with a finite or infinite number of vertices (like gravity in the weak field approximation),
in $d$ space-time dimensions.
Once we have a proof of the SZM identity also for all the two-loop 1PI Green's functions,  
then it would be possible to obtain the pressure of interacting thermal particles in a background of static fields.
This problem is currently under investigation.

\acknowledgments

F. T. Brandt and J. B. Siqueira would like to thank 
CNPq for financial support and Prof. Josif Frenkel for helpful discussions.

\appendix*
\begin{widetext}
\section{}

\subsection{Third topology}

In this appendix we will present the analysis of the three remaining topologies which contribute to the two-point function. 
We begin with the topology shown in Fig.~\ref{fig:top55} which in the static limit yields
\begin{figure}
\begin{center}
\includegraphics[scale=0.45]{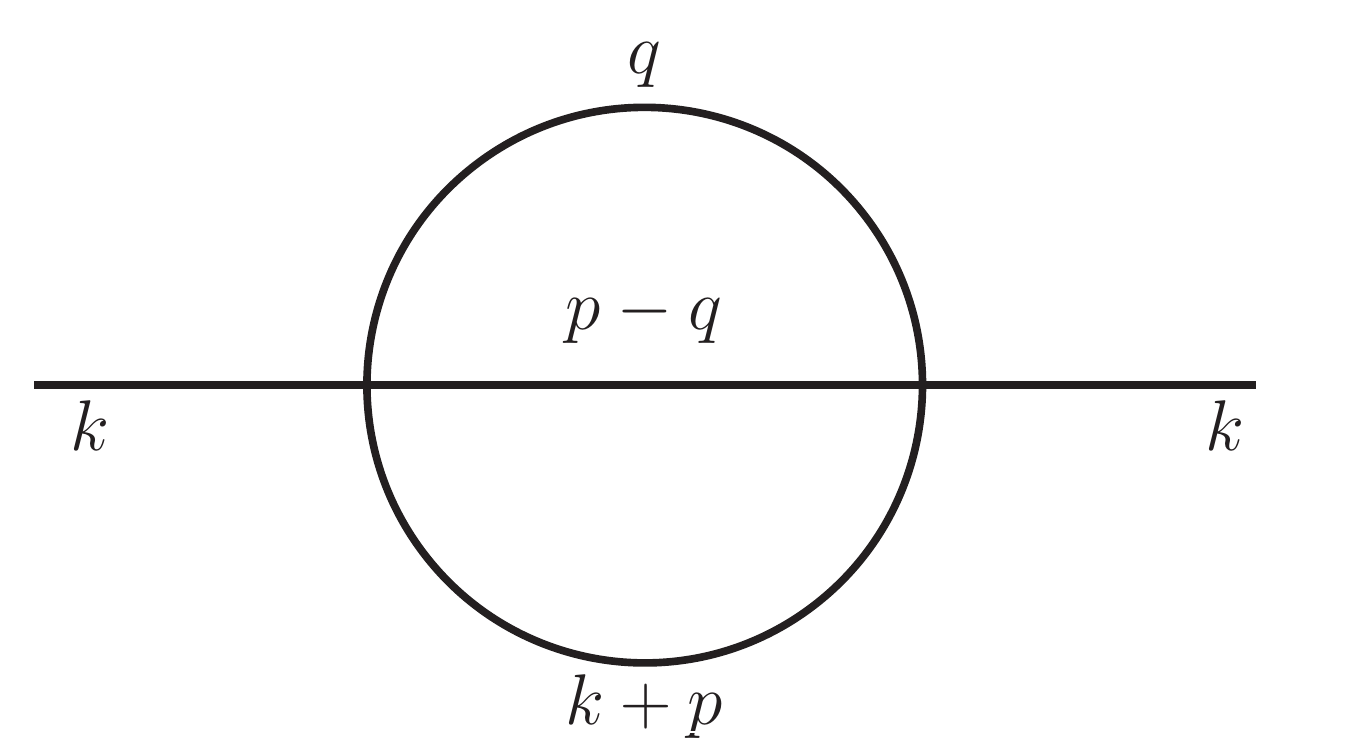}
\end{center}
\caption{Third topology for two-loop self-energy}
\label{fig:top55}
\end{figure}
\begin{align}
\Pi_e^C = \oint_C \frac{dp_0}{2 \pi i} \oint_C \frac{dq_0}{2\pi i} \frac{f^{C}(p_0,q_0)}{ (p_0^2 - w^2) [(p_0-q_0)^2-z^2] (q_0^2- y^2)},
\end{align}
with the notation $w=|\vec{p}+\vec{k}|$, $y=|\vec{q}|$ and $z=|\vec{p}-\vec{q}|$.

Performing the integration in $p_0$, with the help of the residue theorem, we obtain
\begin{align}
\Pi_e^C &= - \oint_C \frac{dq_0}{2\pi i} \left\{\left[  \frac{f^{C}(w,q_0)}{ 2w [(w-q_0)^2-z^2] (q_0^2- y^2)} \right.\right. 
\nonumber \\ &+ \left.\left. \frac{f^{C}(q_0+z,q_0)}{ 2z [(q_0+z)^2-w^2] (q_0^2- y^2)} \right] + [w,z \rightarrow -w,-z] \right\} .
\end{align}  
Similarly, the integration in $q_0$ yields
\begin{align}
\Pi_e^C &= \left\{ \left[  \frac{f^{C}(w,y)}{ 4w y [(w-y)^2-z^2] } + \frac{f^{C}(y+z,y)}{ 4zy [(y+z)^2-w^2]} \right. \right. \nonumber \\ &+ \left. (y \rightarrow -y) \right] + \left[  \frac{f^{C}(w,w+z)}{ 4w z [(w+z)^2- y^2]} + (z \rightarrow -z) \right] \nonumber \\
 &+ \left. \left[ \frac{f^{C}(w,w-z)}{ 4zw [(w-z)^2- y^2]} + (w \rightarrow -w) \right] \right\} \nonumber \\ &+\{w,z \rightarrow -w -z\}.
\end{align}
In the high-temperature limit we consider $w=x+\epsilon$, with $x=|\vec{p}|$, so that the dominant term is
\begin{align}
\Pi_e^C &\simeq \left\{ \left[ \frac{f^{C}(x,y)}{ 4x y [(x-y)^2-z^2] } + \frac{f^{C}(y+z,y)}{ 4zy [(y+z)^2-x^2]} \right. \right. \nonumber \\ &+ \left. (y \rightarrow -y) \right] + \left[  \frac{f^{C}(x,x+z)}{ 4x z [(x+z)^2- y^2]} + (z \rightarrow -z) \right] \nonumber \\
 &+ \left. \left[ \frac{f^{C}(x,x-z)}{ 4zx [(x-z)^2- y^2]} + (x \rightarrow -x) \right] \right\} \nonumber \\ &+\{x,z \rightarrow -x -z\} \nonumber \\
&= \oint_C \frac{dp_0}{2 \pi i} \oint_C \frac{dq_0}{2\pi i} \frac{f^{C}(p_0,q_0)}{ (p_0^2 - x^2) [(p_0-q_0)^2-z^2] (q_0^2- y^2)} \nonumber \\
&= \Pi_0^C,
\end{align}
where we identify the zero-momentum limit of the amplitude
\begin{align}
\Pi_0^C &= \oint_C \frac{dp_0}{2 \pi i} \oint_C \frac{dq_0}{2\pi i} \frac{f^{C}(p_0,q_0)}{ (p_0^2 - x^2) [(p_0-q_0)^2-z^2] (q_0^2- y^2)},
\end{align}
which establishes the SZM identity for this amplitude.

\subsection{Fourth topology}

Let us now consider the topology shown in Fig. \ref{fig:top33} which, in the static limit, has the following form
\begin{figure}[t!]
\begin{center}
\includegraphics[scale=0.45]{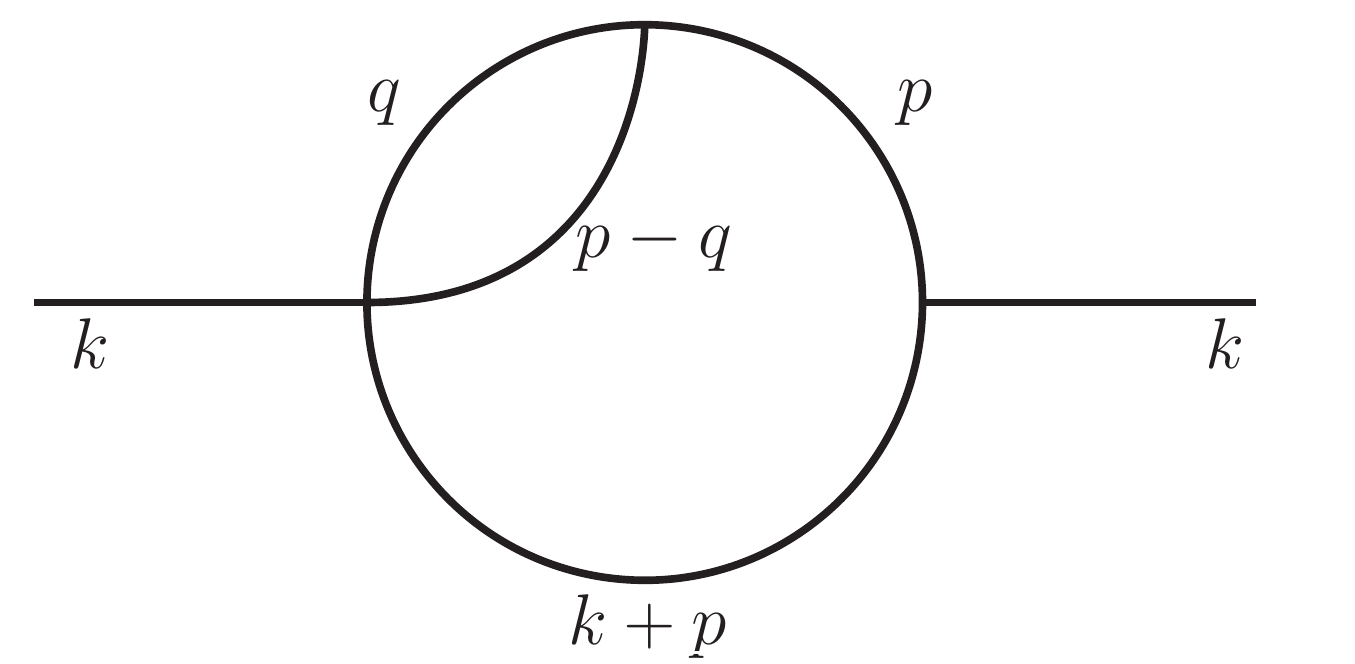}
\end{center}
\caption{Fourth topology for two-loop self-energy}
\label{fig:top33}
\end{figure}
\begin{equation}
\Pi_e^D = \oint_C \frac{dp_0}{2 \pi i} \oint_C \frac{dq_0}{2\pi i} \frac{f^{D}(p_0,q_0)}{ (p_0^2-x^2) (p_0^2 - w^2) [(p_0-q_0)^2-z^2] (q_0^2- y^2)},
\end{equation}
where we are using the notation $x=\left|\vec{p}\right|$, $y=\left|\vec{q}\right|$, $w=|\vec{p}+\vec{k}|$ and $z=\left| \vec{p}-\vec{q} \right|$.

As before, we employ the residue theorem to perform the integration in $p_0$ and in $q_0$, obtaining
\begin{align}
\Pi_e^D &= \left\{ \left[ \frac{f^{D}(x,y)}{4xy(x^2-w^2)[(x-y)^2-z^2]} +  \frac{f^{D}(w,y)}{4wy(w^2-x^2)[(w-y)^2-w^2]} +  \frac{f^{D}(y+z,y)}{4zy[(y+z)^2-x^2][(y+z)^2-w^2]} \right. \right. \nonumber \\ &+ \left. (y \rightarrow -y) \right] +
\left[ \frac{f^{D}(x,x+z)}{4zx(x^2-w^2)[(x+z)^2-y^2]} +  \frac{f^{D}(w,w+z)}{4zw(w^2-x^2)[(w+z)^2-y^2]}+  (z \rightarrow -z) \right] \nonumber \\ 
&+   \left[ \frac{f^{D}(x,x-z)}{4xz(x^2-w^2)[(x-z)^2-y^2]} + (x \rightarrow -x)\right]+ \left.\left[ \frac{f^{D}(w,w-z)}{4wz(w^2-x^2)[(w-z)^2-y^2]} + (w \rightarrow -w) \right] \right\} \nonumber \\ &+ \{ (x,w,z) \rightarrow (-x,-w,-z) \}.
\end{align}


In the high-temperature limit, the relation in Eq. \eqref{eq:aproox00} yields

\begin{align}
\Pi_e^D &\simeq \left\{ \left[ \frac{1}{2x} \frac{\partial}{\partial x} \frac{f^{D}(x,y)}{4xy[(x-y)^2-z^2]}+   \frac{f^{D}(y+z,y)}{4zy[(y+z)^2-x^2]^2} + (y \rightarrow -y) \right] +  
\left[ \frac{1}{2x} \frac{\partial}{\partial x} \frac{f^{D}(x,x+z)}{4zx[(x+z)^2-y^2]} +(z \rightarrow -z) \right] \right. \nonumber \\ 
&+ \left.  \left[ \frac{1}{2x} \frac{\partial}{\partial x} \frac{f^{D}(x,x-z)}{4xz[(x-z)^2-y^2]} + (x \rightarrow -x)\right] \right\} + \{ (x,z) \rightarrow (-x,-z) \}.
\label{eq:pie33}
\end{align}
On the other hand, the zero external four-momentum amplitude is
\begin{equation}
\Pi_0^D = \oint_C \frac{dp_0}{2 \pi i} \oint_C \frac{dq_0}{2\pi i} \frac{f^{D}(p_0,q_0)}{ (p_0^2-x^2)^2 [(p_0-q_0)^2-z^2] (q_0^2- y^2)}.
\end{equation}
\end{widetext}
Using the residue theorem, the integration in $p_0$ gives
\begin{align}
 \Pi_0^D &= - \oint_C \frac{dq_0}{2\pi i}
\left\{\left[\frac{f^{D}(q_0+z,q_0)}{2z [(q_0+z)^2-x^2]^2(q_0^2-y^2)}\right. \right. \nonumber \\ 
&+ \left. \frac{1}{2x} \frac{\partial}{\partial x} \frac{f^{D}(x,q_0)}{2x[(x-q_0)^2-z^2](q_0^2-y^2)} \right] \nonumber \\ 
&+ \left.\left[ (x,z) \rightarrow (-x,-z)\right] \right\}.
\end{align}
Similarly, the integration in $q_0$ produces   
\begin{align}
\Pi_0^D &=\left\{ \left[ \frac{f^{D}(y+z,y)}{4zy[(y+z)^2-x^2]^2} \right.\right. \nonumber \\ &+ \left. \frac{1}{2x} \frac{\partial}{\partial x} \frac{f^{D}(x,y)}{4xy[(x-y)^2-z^2]} + (y \rightarrow -y)  \right]  \nonumber \\
&+ \left[\frac{1}{2x} \frac{\partial}{\partial x} \frac{f^{D}(x,x+z)}{4xz[(x+z)^2-y^2]} +(z \rightarrow -z)\right] \nonumber \\
&+ \left. \left[\frac{1}{2x} \frac{\partial}{\partial x} \frac{f^{D}(x,x-z)}{4zx[(x-z)^2-y^2]} +(x \rightarrow -x) \right] \right\} \nonumber \\  &+ \{(x,z) \rightarrow (-x,-z)\}.
\label{eq:pi033}
\end{align} 
Comparing the static limit [\eqref{eq:pie33}] with the zero momentum limit [\eqref{eq:pi033}] we can verify the SZM  identity for the amplitude shown in Fig.~\ref{fig:top33}.

\subsection{Fifth topology}

\begin{figure}
\begin{center}
\includegraphics[scale=0.45]{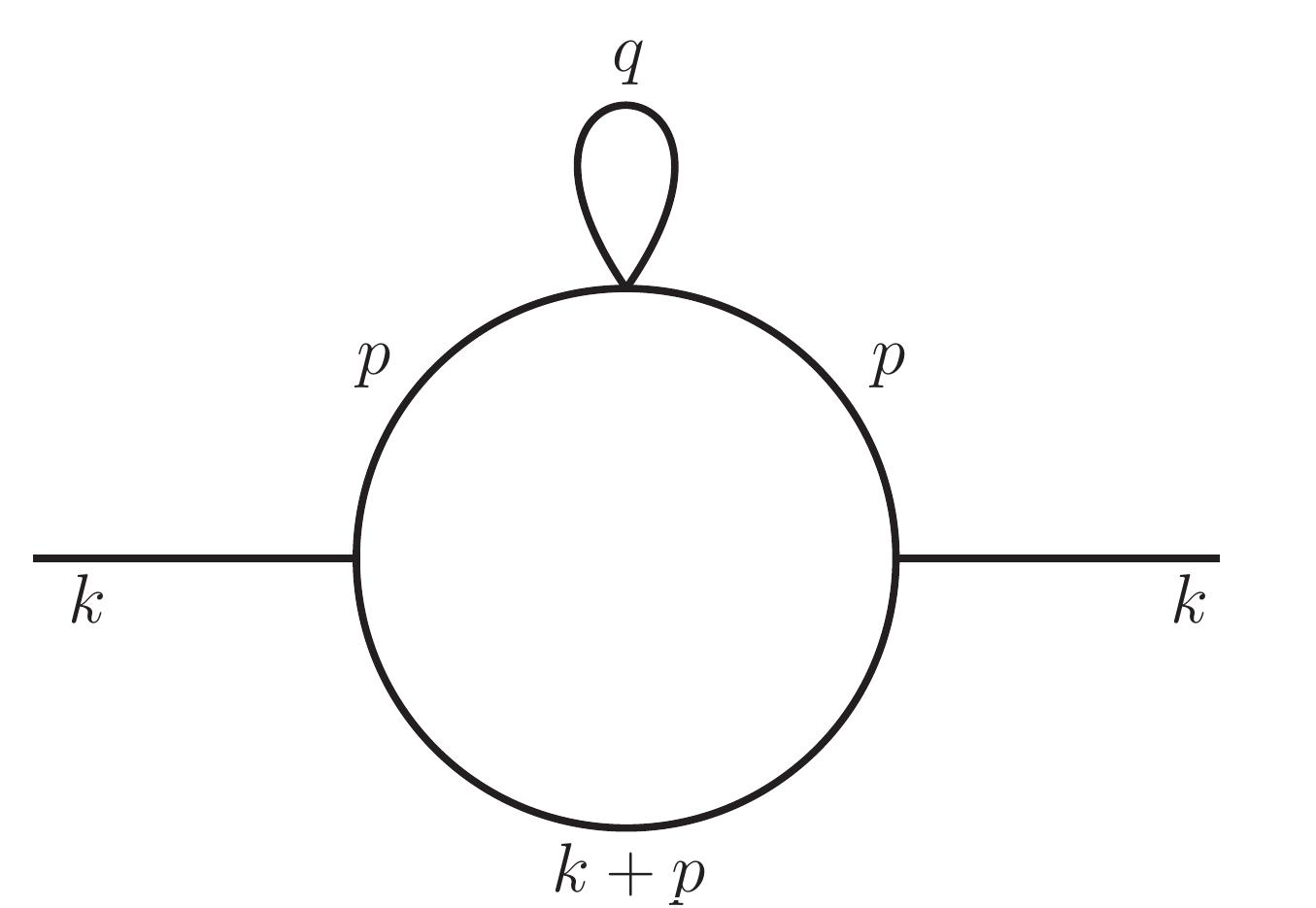}
\end{center}
\caption{Fifth topology for two loops self energy.}
\label{fig:top4}
\end{figure}

The fifth topology for the two-loop self-energy  is shown in
Fig.~\ref{fig:top4}. In the static limit it reduces to 
\begin{equation}
\Pi_e^E = \oint_C \frac {dp_0}{2\pi i} \oint_C \frac {dq_0}{2\pi i}  \frac{f^{E}(p_0,q_0)}{ (p_0^2-x^2)^2(p_0^2-w^2)(q_0^2-y^2)},
\end{equation}
with $x=\left|\vec{p}\right|$, $y=\left|\vec{q}\right|$ and $w=|\vec{p}+\vec{k}|$.
Performing the integration in $p_0$ with the help of the residue theorem, we obtain
\begin{align}
\Pi_e^E &= - \oint_C \frac {dq_0}{2\pi i} \left\{ \frac{f^{E}(w,q_0)}{ 2w (w^2-x^2)^2(q_0^2-y^2)} \right. \nonumber \\ &+ \left. (w \rightarrow -w) \right. \nonumber \\ &+ \left. \frac{1}{2x} \frac{\partial}{\partial x}\frac{f^{E}(x,q_0)}{ 2x(x^2-w^2)(q_0^2-y^2)}\right. \nonumber \\ &+ \left. (x \rightarrow -x)\right\}.
\label{eq:Jjdo}
\end{align}
In this amplitude the internal loops do not overlap, and therefore it is not necessary, for the present purpose, to perform the $q_0$ integration.

In the high-temperature limit we can use the approximation in Eq. \eqref{eq:htldob} and the identity in Eq. \eqref{eq:iiddjjbb} which lead to 
\begin{align}
\Pi_e^E &\simeq -\oint_C \frac {dq_0}{2\pi i} \left\{ \frac{3}{16 x^5} \frac{f^{E}(x,q_0)}{(q_0^2-y^2)}\right. \nonumber \\  &- \left. \frac{3}{16 x^4} \frac{\partial}{\partial x}  \frac{f^{E}(x,q_0)}{(q_0^2-y^2)} +\frac{1}{16 x^3} \frac{\partial^2}{\partial x^2}  \frac{f^{E}(x,q_0)}{(q_0^2-y^2)} \right. \nonumber \\ &+ \left.(x \rightarrow -x )\right\}.
\label{eq:pieu}
\end{align}

When all the external four-momentum vanish, the amplitude reduces to
\begin{equation}
\Pi_0^E = \oint_C \frac {dp_0}{2\pi i} \oint_C \frac {dq_0}{2\pi i}  \frac{f^{E}(p_0,q_0)}{ (p_0^2-x^2)^3(q_0^2-y^2)}.
\end{equation}
Using the residue theorem in order to perform the $p_0$ integration, we obtain
\begin{align}
\Pi_0^E &= -\oint_C \frac {dq_0}{2\pi i} \left[ \frac{1}{2} \lim_{p_0=x} \frac{\partial^2}{\partial p_0^2} \frac{f^{E}(p_0,q_0)}{ (p_0+x)^3(q_0^2-y^2)} \right. \nonumber \\  &+ \left. (x \rightarrow -x) \right] .
\end{align}
Finally, computing the limit, we obtain
\begin{align}
\Pi_0^E &= -\oint_C \frac {dq_0}{2\pi i} \left\{ \frac{3}{16 x^5} \frac{f^{E}(x,q_0)}{(q_0^2-y^2)} - \frac{3}{16 x^4} \frac{\partial}{\partial x}  \frac{f^{E}(x,q_0)}{(q_0^2-y^2)} \right. \nonumber \\ &+ \left. \frac{1}{16 x^3} \frac{\partial^2}{\partial x^2}  \frac{f^{E}(x,q_0)}{(q_0^2-y^2)} +(x \rightarrow -x )\right\}.
\label{eq:pi0u}
\end{align}
Therefore, the static result [\eqref{eq:pieu}] together with  the zero-momentum result [\eqref{eq:pi0u}] implies the SZM identity for the topology shown in Fig.~\ref{fig:top4}.


\begin{thebibliography}{11}
\expandafter\ifx\csname natexlab\endcsname\relax\def\natexlab#1{#1}\fi
\expandafter\ifx\csname bibnamefont\endcsname\relax
  \def\bibnamefont#1{#1}\fi
\expandafter\ifx\csname bibfnamefont\endcsname\relax
  \def\bibfnamefont#1{#1}\fi
\expandafter\ifx\csname citenamefont\endcsname\relax
  \def\citenamefont#1{#1}\fi
\expandafter\ifx\csname url\endcsname\relax
  \def\url#1{\texttt{#1}}\fi
\expandafter\ifx\csname urlprefix\endcsname\relax\def\urlprefix{URL }\fi
\providecommand{\bibinfo}[2]{#2}
\providecommand{\eprint}[2][]{\url{#2}}

\bibitem[{\citenamefont{Braaten and
  Pisarski}(1990{\natexlab{a}})}]{Braaten:1990mz}
\bibinfo{author}{\bibfnamefont{E.}~\bibnamefont{Braaten}} \bibnamefont{and}
  \bibinfo{author}{\bibfnamefont{R.~D.} \bibnamefont{Pisarski}},
  \bibinfo{journal}{Nucl. Phys.} \textbf{\bibinfo{volume}{B337}},
  \bibinfo{pages}{569} (\bibinfo{year}{1990}{\natexlab{a}});
%
\textbf{\bibinfo{volume}{B339}},
  \bibinfo{pages}{310} (\bibinfo{year}{1990}{\natexlab{b}}).

\bibitem[{\citenamefont{Frenkel and Taylor}(1990)}]{Frenkel:1990br}
\bibinfo{author}{\bibfnamefont{J.}~\bibnamefont{Frenkel}} \bibnamefont{and}
  \bibinfo{author}{\bibfnamefont{J.~C.} \bibnamefont{Taylor}},
  \bibinfo{journal}{Nucl. Phys.} \textbf{\bibinfo{volume}{B334}},
  \bibinfo{pages}{199} (\bibinfo{year}{1990});
%
\textbf{\bibinfo{volume}{B374}},
 \bibinfo{pages}{156} (\bibinfo{year}{1992}).

\bibitem[{\citenamefont{Kapusta}(1989)}]{kapusta:book89}
\bibinfo{author}{\bibfnamefont{J.~I.} \bibnamefont{Kapusta}},
  \emph{\bibinfo{title}{Finite Temperature Field Theory}}
  (\bibinfo{publisher}{Cambridge University Press},
  \bibinfo{address}{Cambridge, England}, \bibinfo{year}{1989}).

\bibitem[{\citenamefont{Bellac}(1996)}]{lebellac:book96}
\bibinfo{author}{\bibfnamefont{M.~L.} \bibnamefont{Bellac}},
  \emph{\bibinfo{title}{Thermal Field Theory}} (\bibinfo{publisher}{Cambridge
  University Press}, \bibinfo{address}{Cambridge, England},
  \bibinfo{year}{1996}).

\bibitem[{\citenamefont{Brandt et~al.}(2009)\citenamefont{Brandt, Frenkel,
  McKeon, and Siqueira}}]{Brandt:2009rq}
\bibinfo{author}{\bibfnamefont{F.~T.} \bibnamefont{Brandt}},
  \bibinfo{author}{\bibfnamefont{J.}~\bibnamefont{Frenkel}},
  \bibinfo{author}{\bibfnamefont{D.~G.~C.} \bibnamefont{McKeon}},
  \bibnamefont{and} \bibinfo{author}{\bibfnamefont{J.~B.}
  \bibnamefont{Siqueira}}, \bibinfo{journal}{Phys. Rev.}
  D~\textbf{\bibinfo{volume}{80}}, \bibinfo{pages}{025024}
  (\bibinfo{year}{2009}).

\bibitem[{\citenamefont{Brandt et~al.}(2006)\citenamefont{Brandt, Frenkel, and
  Muramoto}}]{Brandt:2006bf}
\bibinfo{author}{\bibfnamefont{F.~T.} \bibnamefont{Brandt}},
  \bibinfo{author}{\bibfnamefont{J.}~\bibnamefont{Frenkel}}, \bibnamefont{and}
  \bibinfo{author}{\bibfnamefont{C.}~\bibnamefont{Muramoto}},
  \bibinfo{journal}{Nucl. Phys.} \textbf{\bibinfo{volume}{B754}},
  \bibinfo{pages}{146} (\bibinfo{year}{2006}).

\bibitem[{\citenamefont{Brandt and Frenkel}(1993)}]{Brandt:1993dk}
\bibinfo{author}{\bibfnamefont{F.~T.} \bibnamefont{Brandt}} \bibnamefont{and}
  \bibinfo{author}{\bibfnamefont{J.}~\bibnamefont{Frenkel}},
  \bibinfo{journal}{Phys. Rev.} D~\textbf{\bibinfo{volume}{47}},
  \bibinfo{pages}{4688} (\bibinfo{year}{1993}).

\bibitem[{\citenamefont{Brandt et~al.}(1995)\citenamefont{Brandt, Frenkel, and
  Taylor}}]{Brandt:1995mv}
\bibinfo{author}{\bibfnamefont{F.~T.} \bibnamefont{Brandt}},
  \bibinfo{author}{\bibfnamefont{J.}~\bibnamefont{Frenkel}}, \bibnamefont{and}
  \bibinfo{author}{\bibfnamefont{J.~C.} \bibnamefont{Taylor}},
  \bibinfo{journal}{Nucl. Phys.} \textbf{\bibinfo{volume}{B437}},
  \bibinfo{pages}{433} (\bibinfo{year}{1995}).

\bibitem[{\citenamefont{Brandt and Siqueira}(2012)}]{Brandt:2012ei}
\bibinfo{author}{\bibfnamefont{F.~T.} \bibnamefont{Brandt}} \bibnamefont{and}
  \bibinfo{author}{\bibfnamefont{J.~B.} \bibnamefont{Siqueira}},
  \bibinfo{journal}{Phys. Rev.} D~\textbf{\bibinfo{volume}{85}},
  \bibinfo{pages}{067701} (\bibinfo{year}{2012}).

\bibitem[{\citenamefont{Frenkel et~al.}(2009)\citenamefont{Frenkel, Pereira,
  and Takahashi}}]{Frenkel:2009pi}
\bibinfo{author}{\bibfnamefont{J.}~\bibnamefont{Frenkel}},
  \bibinfo{author}{\bibfnamefont{S.~H.} \bibnamefont{Pereira}},
  \bibnamefont{and}
  \bibinfo{author}{\bibfnamefont{N.}~\bibnamefont{Takahashi}},
  \bibinfo{journal}{Phys. Rev.} D~\textbf{\bibinfo{volume}{79}},
  \bibinfo{pages}{085001} (\bibinfo{year}{2009}).

\bibitem{Brandt:2009ht} 
  F.~T.~Brandt, J.~Frenkel and J.~C.~Taylor,
  Nucl.\ Phys.\ B {814}, 366 (2009).

\bibitem{Mottola:2009mi} 
  E.~Mottola and Z.~Szep,
  Phys.\ Rev.\ D {\bf 81}, 025014 (2010).

\end{thebibliography}

\end{document}